\documentclass{ieeeaccess}

\usepackage{cite}
\usepackage{amsmath,amssymb,amsfonts}
\usepackage{algorithmic}
\usepackage{graphicx}
\usepackage{textcomp}

\usepackage{subfigure}
\usepackage{dsfont}

\usepackage{colortbl}

\usepackage[export]{adjustbox}

\usepackage{placeins}
\usepackage[normalem]{ulem}

\usepackage{multirow}
\usepackage{makecell}
\usepackage{booktabs}

\usepackage{float}

\usepackage{pgfplots}
\usepackage{pgfplotstable}

\usepackage{tikz}
\NewSpotColorSpace{PANTONE}
\AddSpotColor{PANTONE} {PANTONE3015C} {PANTONE\SpotSpace 3015\SpotSpace C} {1 0.3 0 0.2}
\SetPageColorSpace{PANTONE}

\pgfplotsset{width=10cm,compat=1.15}
\usepgfplotslibrary{statistics}
\usetikzlibrary{pgfplots.groupplots}

\usepackage{url}

\usepackage[hidelinks]{hyperref}
\hypersetup{
    colorlinks=true,
    linkcolor=black,
    filecolor=black,
    urlcolor=gray,
    citecolor=black,
    breaklinks=true
}

\definecolor{mylightgray}{gray}{0.9}

\newcommand{\bbox}{\protect\raisebox{1pt}{\protect\tikz \protect\draw[black,fill=black] (1,1) circle (0.5ex);}}
\newcommand{\wbox}{\protect\raisebox{1pt}{\protect\tikz \protect\draw[black,fill=white] (1,1) circle (0.5ex);}}

\definecolor{c1}{RGB}{ 246.2025  131.8860    8.7975}    % orange
\definecolor{c2}{RGB}{   128.0100  128.0100  128.0100}  % gray
\definecolor{c3}{RGB}{          0  202.2405  255.0000}  % cyan
\definecolor{c4}{RGB}{          0         0  255.0000}  % blue
\definecolor{c5}{RGB}{          0  255.0000         0}  % green
\definecolor{c6}{RGB}{          0   87.9240         0}  % dark green

\begin{document}

\twocolumn[
\begin{@twocolumnfalse}
% Author’s pre-print (ie pre-refereeing) /post-print (ie final draft post-refereeing) accepted for publication/published in the IEEE Transactions on XXX (Jan. 2021).\\
Authors’ post-print accepted for publication/published in the IEEE Access (Mar. 2022).\\

\copyright 2022 IEEE. Personal use of this material is permitted. Permission from IEEE must be obtained for all other users, including reprinting/ republishing this material for advertising or promotional purposes, creating new collective works for resale or redistribution to servers or lists, or reuse of any copyrighted components of this work in other works.\\
DOI: \url{https://doi.org/10.1109/ACCESS.2022.3166906}\\ %fill this for post-print
URL: \url{https://ieeexplore.ieee.org/document/9756023}\\ %fill this for post-print

% Cite: J. Doe, A. Einstein, “Title of the article,” \textit{IEEE Transactions on xxx}, vol. xx, no. xx, id. xx, Jan. 2021. \\ %fill this for post-print
Cite: A. Xompero, S. Donaher, V. Iashin, F. Palermo, G. Solak, C. Coppola, R. Ishikawa, Y. Nagao, R. Hachiuma, Q. Liu, F. Feng, C. Lan, R. H. M. Chan, G. Christmann, J. Song, G. Neeharika, C. K. T. Reddy, D. Jain, B. U. Rehman, A. Cavallaro, "The CORSMAL benchmark for the prediction of the properties of containers," in \textit{IEEE Access}, Early Access.

\end{@twocolumnfalse}
]

\clearpage

\history{Received March 11, accepted March 26, 2022. Date of publication xxxx 00, 0000, date of current version xxxx 00, 0000.}
\doi{10.1109/ACCESS.2022.3166906}

\title{The CORSMAL benchmark for the prediction of the properties of containers}
\author{\uppercase{Alessio Xompero}\authorrefmark{1}, 
        \uppercase{Santiago Donaher}\authorrefmark{1},
        \uppercase{Vladimir Iashin}\authorrefmark{2},
        \uppercase{Francesca Palermo}\authorrefmark{1},
        \uppercase{Gökhan Solak}\authorrefmark{1},
        \uppercase{Claudio Coppola}\authorrefmark{1},
        \uppercase{Reina Ishikawa}\authorrefmark{3},
        \uppercase{Yuichi Nagao}\authorrefmark{3},
        \uppercase{Ryo Hachiuma}\authorrefmark{3},
        \uppercase{Qi Liu}\authorrefmark{4},
        \uppercase{Fan Feng}\authorrefmark{4}, 
        \uppercase{Chuanlin Lan}\authorrefmark{4},
        \uppercase{Rosa H.~M. Chan}\authorrefmark{4},
        \uppercase{Guilherme Christmann}\authorrefmark{5},
        \uppercase{Jyun-Ting Song}\authorrefmark{5},
        \uppercase{Gonuguntla Neeharika}\authorrefmark{6}, 
        \uppercase{Chinnakotla K.~T. Reddy}\authorrefmark{6},
        \uppercase{Dinesh Jain}\authorrefmark{7},
        \uppercase{Bakhtawar Ur Rehman}\authorrefmark{8},
        \uppercase{Andrea Cavallaro}\authorrefmark{1}%
        }
\address[1]{Centre for Intelligent Sensing, Queen Mary University of London, London E1 4NS, U.K.}
\address[2]{Tampere University, 33100 Tampere, Finland}
\address[3]{Keio University, Kanagawa 223-8522, Japan}
\address[4]{City University of Hong Kong, Hong Kong}
\address[5]{National Taiwan Normal University, Taipei 106, Taiwan}
\address[6]{IIT~Bhubaneswar, Bhubaneswar 751013, India}
\address[7]{IIT~Hyderabad, Hyderabad 502285, India}
\address[8]{Islamabad 44000, Pakistan}
\tfootnote{This work was supported by the CHIST-ERA program through the Project CORSMAL under U.K. Engineering and Physical Sciences Research Council (EPSRC) under Grant EP/S031715/1. This work involved human subjects or animals in its research. Approval of all ethical and experimental procedures and protocols was granted by the Queen Mary Ethics of Research Committee under Application No. QMREC2344a.}

\markboth
{Xompero \headeretal: CORSMAL Benchmark for Prediction of Properties of Containers}
{Xompero \headeretal: CORSMAL Benchmark for Prediction of Properties of Containers}

\corresp{Corresponding author: Alessio Xompero (e-mail: a.xompero@qmul.ac.uk).}

\begin{abstract}
The contactless estimation of the weight of a container and the amount of its content manipulated by a person are key pre-requisites for safe human-to-robot handovers. However, opaqueness and transparencies of the container and the content, and variability of materials, shapes, and sizes, make this estimation difficult. In this paper, we present a range of methods and an open framework to benchmark acoustic and visual perception for the estimation of the capacity of a container, and the type, mass, and amount of its content. The framework includes a dataset, specific tasks and performance measures. We conduct an in-depth comparative analysis of methods that used this framework and audio-only or vision-only baselines designed from related works. Based on this analysis, we can conclude that audio-only and audio-visual classifiers are suitable for the estimation of the type and amount of the content using different types of convolutional neural networks, combined with either recurrent neural networks or a majority voting strategy, whereas computer vision methods are suitable to determine the capacity of the container using regression and geometric approaches. Classifying the content type and level using only audio achieves a weighted average F1-score up to 81\% and 97\%, respectively. Estimating the container capacity with vision-only approaches and estimating the filling mass with audio-visual multi-stage approaches reach up to 65\% weighted average capacity and mass scores. These results show that there is still room for improvement on the design of new methods. These new methods can be ranked and compared on the individual leaderboards provided by our open framework.
\end{abstract}

\begin{keywords}
Acoustic signal processing, image and video signal processing, audio-visual classification, object properties recognition
\end{keywords}

\titlepgskip=-15pt

\maketitle

\section{Introduction}
\label{sec:introduction}

\PARstart{P}{eople} interact daily with household containers, such as cups, drinking glasses, mugs, bottles, and food boxes. Methods to estimate the physical properties (e.g.,~weight and shape) of these containers could support human-robot cooperation~\cite{Sanchez-Matilla2020,Medina2016,Rosenberger2021RAL,Ortenzi2021TRO,Yang2021ICRA}, video annotation and captioning. Methods should generalize to unknown container instances and operate with only limited prior knowledge, such as generic categories of containers and contents~\cite{Liang2020MultimodalPouring,Sanchez-Matilla2020,Modas2021ArXiv}. However, the material, texture, transparency, and shape vary considerably across containers and may change with the content. Furthermore, the content may not be visible due to the opaqueness of the container or because of hand occlusions. For these reasons, predicting the physical properties of containers is a challenging task. The combination of sensing modalities, namely RGB images, depth, and audio, may help to overcome challenges such as noisy scenarios, already filled containers with absence of sound, occlusions, or transparent objects whose depth data may be highly inaccurate~\cite{Xompero2020ICASSP_LoDE}.

\begin{figure}[h!]
    \centering
    \includegraphics[width=\columnwidth]{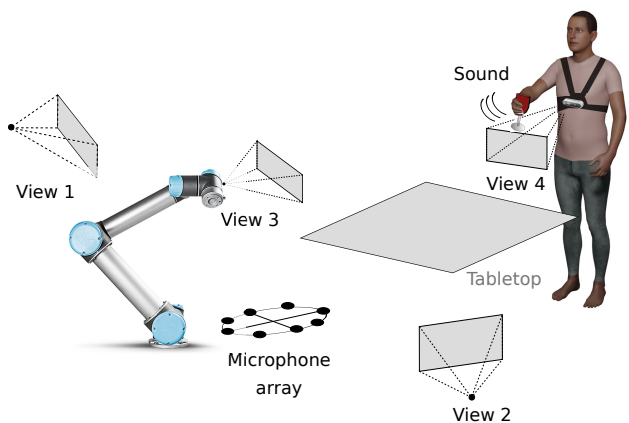}
    \caption{The multi-modal, multi-sensor system used to record a person manipulating a container and its content. The system includes two third-person view cameras (at the two sides of the robot), a first-person view camera mounted on the robot, a first-person view from the body-worn camera on the person and a 8-microphone circular array (placed next to the robot arm).}
%    \vspace{-10pt}
    \label{fig:setup}
\end{figure}

The contributions of this paper include:
\begin{itemize}
    \item A novel framework for the comparison of methods that estimate the physical properties of containers and their content, when a person manipulates the container (see~Fig.~\ref{fig:setup});
    \item The definition of three tasks, such as the classification of the content amount, the classification of the content type, and the estimation of the container capacity, and related performances measures, including the indirect filling mass estimation based on the three tasks, for the framework;
    \item The design of 12 audio-only baselines and one vision-only baseline for the tasks of classifying the content level and the content type based on related approaches from the literature;
    \item A formal review, a comparative analysis, and an in-depth discussion of methods that used the framework to address this problem;
    \item The results of an international benchmarking challenge\footnote{\url{https://corsmal.eecs.qmul.ac.uk/challenge2020.html}}.
\end{itemize}

The paper is organized as follows. Section~\ref{sec:relatedwork} discusses related works. Section~\ref{sec:framework} presents the benchmarking framework, including a multi-modal dataset, tasks for the estimation of the container and content properties, and corresponding performance measures. Section~\ref{sec:fillingclass}  reviews the methods that used the framework for the tasks of filling type and level classification.
Section~\ref{sec:capacity} reviews the methods that used the framework for the task of container capacity estimation. Section~\ref{sec:analysis} discusses and compares the results of the methods under analysis. Section~\ref{sec:conclusions} concludes the paper and  discusses future research direction.

\section{Related work}
\label{sec:relatedwork}

In this section, we discuss the object properties that are commonly estimated in the literature. We then review methods that recognize the content type, estimate the amount of content in a container, or estimate the container capacity, based on their approaches and input modalities.   

Most of the works in the literature focus on object recognition, object shape and size reconstruction in 3D, as well as pose estimation of a variety of objects using visual data and objects standing on a surface~\cite{He2017ICCV_MaskRCNN,Wang2019CVPR_NOCS,Yang2020CVPR,objectron2020,chen2020category,Chen2020CVPR_CASS,hodan2018bop,kaskman2019homebreweddb}. Object properties, such as transparency, are often tackled independently with ad-hoc designed approaches for 3D shape reconstruction, object localization in 3D, or 6D pose estimation~\cite{Xompero2020ICASSP_LoDE,Liu2020CVPR_KeyPose,sajjan2020clear,Philips2016RSS}. Recognizing different high-level properties, such as the type and amount of multiple filling materials, the capacity of the container, and the overall weight of the object (i.e., the container with its content) is not yet well-investigated. 

Recognizing the \textit{content type} within a container is addressed only for general food recognition using visual information~\cite{Bossard2014ECCV,Kawano2014ACMM,Zhao_2021_WACV}. Audio modality is commonly used for the recognition of general environmental sounds using the combination of traditional features and machine learning classifiers -- e.g., k-Nearest Neighbour  kNN~\cite{Cover1967NearestNP}, Support Vector Machine (SVM)~\cite{Cortes1995SupportN}, and Random Forest (RF)~\cite{Breiman2004RandomF} --, or deep learning approaches -- e.g., convolutional neural networks (CNNs)~\cite{Piczak2015ESCDF}. Examples of traditional acoustic features are spectrograms, zero-crossing rate (ZCR), Mel-frequency Cepstrum Coefficients (MFCCs), chromogram, Mel-scaled spectrogram, spectral contrast, and tonal centroid features (tonnetz)~\cite{Vivek2020ICCSP,Shao2008RobustSI,Ghosal2018MusicGR,R2020AudioVS}. However, there are no unimodal or audio-visual approaches that recognize the content type during the manipulation of different containers held by person and together with other physical properties.

\begin{table*}[t!]
    \centering
    \footnotesize
    \setlength\tabcolsep{2pt}
    \caption{Methods that used the CORSMAL framework for filling level, filling type, and container capacity estimation. Methods are evaluated on the CORSMAL Container Manipulation dataset. 
    }
    \begin{tabular}{lccc>{\raggedright\arraybackslash}p{10cm}cccccccc}
    \specialrule{1.2pt}{0.2pt}{1pt}
    \textbf{Ref.} &\textbf{FL} & \textbf{FT} & \textbf{CC} & \multicolumn{1}{c}{\textbf{Description}} &  \textbf{App} &   \textbf{JLT} &  \textbf{L} & \textbf{Gr} &  \textbf{A} & \textbf{R} &  \textbf{D} & \textbf{Temp.} \\
    \specialrule{1.2pt}{0.2pt}{1pt} 
    M1 & \bbox & \bbox & \wbox & STFT + FCNN & C & \wbox & \bbox & \bbox & \bbox & -- & -- & \wbox \\
    \midrule
    \multirow{2}{*}{M2~\cite{Christmann2020NTNU}} & \wbox & \bbox & \wbox & MFCCs + CNN & C & \wbox & \bbox & \bbox & \bbox & -- & -- & \wbox \\
    & \wbox & \wbox & \bbox & CNN with region of interest and bounding box size & R & \wbox & -- & -- & \wbox & 1 & 1 & \wbox \\
    \midrule
    \multirow{2}{*}{M3~\cite{Liu2020ICPR}} & \bbox & \bbox & \wbox & Spectrogram + object-specific MLP selected via majority voting of per-frame object detection across multiple views & C & \wbox & \bbox & \bbox & \bbox & 4&  --  & \wbox  \\
    & \wbox & \wbox & \bbox & Gaussian processes & R & \wbox & -- & -- & \wbox &  4 &  -- & \wbox  \\
    \midrule
    \multirow{3}{*}{M4~\cite{Ishikawa2020ICPR}} & \bbox & \wbox & \wbox & Multi-channel spectrogram + CNN + LSTM & C & \wbox & \bbox & \bbox & \bbox &  -- &  -- & \bbox \\
    & \wbox & \bbox & \wbox & Multi-channel spectrogram + CNN +  majority voting & C & \wbox & \bbox & \bbox & \bbox & --  & --  & \bbox \\
    & \wbox & \wbox & \bbox & Point cloud + 3D cuboid approximation  & G & \wbox & -- & -- & \wbox & 1& 1 & \bbox \\
    \midrule 
    \multirow{3}{*}{M5~\cite{Iashin2020ICPR}} & \bbox & \wbox & \wbox & R(2+1)D+GRU (video), CNN+GRU (audio), A34F+RF (audio), Late fusion (averaging) & C & \wbox & \bbox & \bbox & \bbox &  1 &  --  & \bbox \\
    & \wbox & \bbox & \wbox & CNN+GRU (audio), A34F+RF (audio), Late fusion (averaging) & C & \wbox & \bbox & \bbox & \bbox &  -- & -- & \bbox \\
    & \wbox & \wbox & \bbox & Energy minimization + 3D cylinder approximation & G & \wbox & -- & -- & \wbox &  2 &  -- & \wbox \\
    \midrule 
    \multirow{3}{*}{M6~\cite{Donaher2021EUSIPCO_ACC}} & \bbox & \bbox & \wbox & Cropped, resized, reshaped spectrogram + kNN/SVM/RF & C & \bbox & \bbox & \bbox & \bbox &  -- &  -- & \wbox \\
    & \bbox & \bbox & \wbox & Cropped and resized spectrogram + CNN                 & C & \bbox & \bbox & \bbox & \bbox &  -- &  --  & \wbox \\
    & \bbox & \bbox & \wbox & Cropped and resized spectrogram + Hierarchy of 3 CNNs & C & \bbox & \bbox & \bbox & \bbox &  -- &  -- & \wbox  \\
    \specialrule{1.2pt}{0.2pt}{5pt}
    \multicolumn{13}{l}{\scriptsize{\parbox{.95\linewidth}{KEY -- FL:~filling level estimation, FT:~filling type estimation, CC:~container capacity estimation, App:~approach, JLT:~joint filling type and level classification, A:~audio, R:~number of used RGB views, D:~number of views used with depth data, L:~liquids, Gr:~granular materials, Temp.:~temporal,  C:~classification, R:~regression, G:~projective geometry, CNN:~convolutional neural network, STFT:~short-term Fourier transform, FCNN:~fully connected neural network, MLP:~multi-layer perceptron, LSTM:~Long-Short Term Memory, GRU:~Gated Recurrent Unit, kNN:~k-Nearest Neighbour, SVM:~support vector machine. RF:~random forest, A34F:~34 audio features~\cite{Giannakopoulos2015PLoS_ONE_pyAudioAnalysis} consisting of zero crossing rate, energy, entropy of energy, spectral centroid, spectral spread, spectral entropy, spectral flux, spectral rolloff, Mel-frequency Cepstrum Coefficients (MFFCs), chroma vector and chroma deviation.}}}\\
\end{tabular}
\label{tab:sotafillinglevel}
% \vspace{-10pt}
\end{table*}

For \textit{content level} estimation, some methods regress or classify the property using CNNs and a single image~\cite{Mottaghi2017ICCV,Modas2021ArXiv}, or use temporal information from sequences of RGB or RGB-D data to track the change in the amount during a {mechanical} action~\cite{Schenck2017RSS,Do2016,Do2018}. Other methods use the sound signals generated by the contact of the content with a container during a manipulation~\cite{griff,Ikeno2015,clarke,Liang2019AudioPouring}. For example, the level of unknown liquids within containers standing on a surface is regressed or classified by using approaches such as Kalman Filter and recurrent neural networks with edge features or spectrograms~\cite{Do2016,Do2018,Liang2019AudioPouring}. 
For the estimation of the \textit{capacity} of a container, one work trained a CNN using an RGB image of one or more containers standing on a surface~\cite{Mottaghi2017ICCV}. However, all of these approaches are often designed and evaluated on scenarios with only standing containers, and with limited variability in
the data.

Unlike previous works, in the next sections we present an open framework for the estimation of multiple physical properties of containers and contents as they are manipulated by a person. We also discuss methods that used this framework based on the modalities used as input, the features extracted, and the type of approach (regression, classification, or geometry-based)~\cite{Christmann2020NTNU,Iashin2020ICPR,Ishikawa2020ICPR,Liu2020ICPR,Donaher2021EUSIPCO_ACC} (see Table~\ref{tab:sotafillinglevel}).

%%%%%%%%%%%%%%%%%%%%%%%%%%%%%%%%%
\section{Benchmarking framework}
\label{sec:framework}

\subsection{Containers, fillings, scenarios}

The dataset includes audio-visual-inertial recordings of people manipulating a range of containers that vary in shape, size, material, transparency, and deformability, and a set of contents under different scenarios with increasing level of difficulty due to the type of occlusions. 

CORSMAL Containers Manipulation~\cite{Xompero_CCM} is a dataset consisting of 1,140 audio-visual recordings with 12 human subjects manipulating 15 containers, split into 5 cups, 5 drinking glasses, and 5 food boxes. These containers are made of different materials, such as plastic, glass, and cardboard. Each container can be empty or filled with water, rice or pasta at two different levels of fullness: 50\% and 90\% with respect to the capacity of the container. The combination of containers and contents results in a total of 95 configurations acquired for three scenarios with an increasing level of difficulty caused by occlusions or subject motions. 

In the first scenario, the subject sits in front of the robot, while a container is on a table. The subject either pours the content into the empty container, while avoiding touching the container, or shakes an already filled food box. Afterwards, the subject initiates the handover of the container to the robot. In the second scenario, the subject sits in front of the robot, while holding a container before starting the manipulation. In the third scenario, a container is held by the subject while standing to the side of the robot, potentially visible only on the third-person camera view. After the manipulation, the subject takes a few steps and initiates the handover of the container in front of the robot. 
Each scenario is recorded with two different backgrounds and under two different lighting conditions. The first background condition involves a plain tabletop with the subject wearing a texture-less t-shirt, while the second background condition involves the table covered with a graphics-printed tablecloth and the subject wearing a patterned shirt. The first lighting condition is based on artificial illumination as provided by lights mounted on the ceiling of the room. The second lighting condition uses two controlled artificial lights placed at the sides of the robot and illuminating the area where the manipulation is happening. 
Each subject executed the 95 configurations for each scenario and for each background/illumination condition\footnote{Ethical approval (QMREC2344a) obtained at Queen Mary University of London. Consent from the subjects was collected before data collection.}.

%%%%%%%%%%%%%%%%%%%%%%%%%%%%%%%%%%%%%%%%%%%%%%%%%%%%%%%%%%%%%%5555
\definecolor{ts1}{RGB}{127,127,127}    % Empty
\definecolor{ts2}{RGB}{245,130,48}  % P5: orange
\definecolor{ts3}{RGB}{255,255,25}  % P9: yellow
\definecolor{ts4}{RGB}{170,110,40}  % R5: brown
\definecolor{ts5}{RGB}{70,240,240}  % W5: cyan
\definecolor{ts6}{RGB}{128,0,0}  % R9: maroon
\definecolor{ts7}{RGB}{0,0,128}  % W9: navy

\pgfplotstableread{masses.txt}\masses
\begin{figure*}[h!]
    \centering
    \begin{tikzpicture}
    \begin{axis}[
    axis x line*=bottom,
    axis y line*=left,
    enlarge x limits=false,
    ybar,
    width=\linewidth,
	bar width=3pt,
    xmin=0.5,xmax=9.5,
    xtick=data,
    height=0.75\columnwidth,
    ymin=0,  ymax=3000,
    ylabel={Mass (g)},
    label style={font=\footnotesize},
    tick label style={font=\footnotesize},
    ymajorgrids=true,
    xticklabels={},
    ]
    \addplot+[ybar, black, fill=white, draw opacity=0.5] table[x=CoID,y=E]{\masses};
    \addplot+[ybar, black, fill=ts2, draw opacity=0.5] table[x=CoID,y=P5]{\masses};
    \addplot+[ybar, black, fill=ts3, draw opacity=0.5] table[x=CoID,y=P9]{\masses};
    \addplot+[ybar, black, fill=gray, draw opacity=0.5] table[x=CoID,y=R5]{\masses};
    \addplot+[ybar, black, fill=ts5, draw opacity=0.5] table[x=CoID,y=W5]{\masses};
    \addplot+[ybar, black, fill=ts6, draw opacity=0.5] table[x=CoID,y=R9]{\masses};
    \addplot+[ybar, black, fill=ts7, draw opacity=0.5] table[x=CoID,y=W9]{\masses};
    \end{axis}
    \node[inner sep=0pt] (flute1) at (0.9,-0.9)
    {\includegraphics[width=.12\columnwidth]{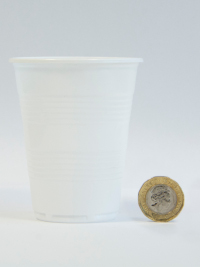}};
    \node[inner sep=0pt] (flute1) at (2.7,-0.9)
    {\includegraphics[width=.12\columnwidth]{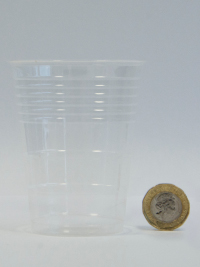}};
    \node[inner sep=0pt] (flute1) at (4.5,-0.9)
    {\includegraphics[width=.12\columnwidth]{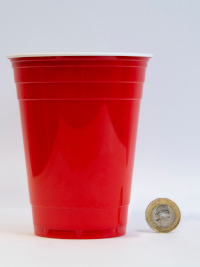}};
    \node[inner sep=0pt] (flute1) at (6.3,-0.9)
    {\includegraphics[width=.12\columnwidth]{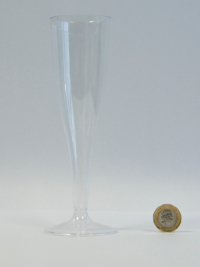}};
    \node[inner sep=0pt] (flute1) at (8.1,-0.9) 
    {\includegraphics[width=.12\columnwidth]{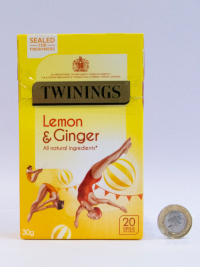}};
    \node[inner sep=0pt] (flute1) at (9.9,-0.9)
    {\includegraphics[width=.12\columnwidth]{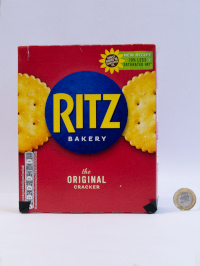}};
    \node[inner sep=0pt] (flute1) at (11.7,-0.9)
    {\includegraphics[width=.12\columnwidth]{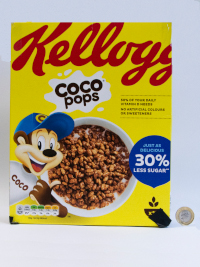}};
    \node[inner sep=0pt] (flute1) at (13.5,-0.9)
    {\includegraphics[width=.12\columnwidth]{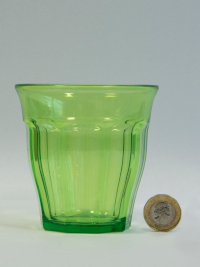}};
    \node[inner sep=0pt] (flute1) at (15.3,-0.9)
    {\includegraphics[width=.12\columnwidth]{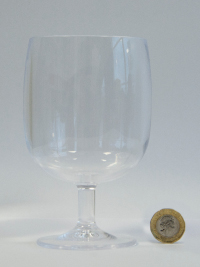}};
    \end{tikzpicture}
    \caption{The mass of objects (container and content) in the training set of the CORSMAL Containers Manipulation dataset. The class \textit{empty} corresponds to the mass of the container, which is known.
    Legend:
    \protect\raisebox{2pt}{\protect\tikz \protect\draw[gray,fill=white] (0,0) rectangle (1.ex,1.ex);}~Empty,
    \protect\raisebox{2pt}{\protect\tikz \protect\draw[gray,fill=ts2] (0,0) rectangle (1.ex,1.ex);}~P5,
    \protect\raisebox{2pt}{\protect\tikz \protect\draw[gray,fill=ts3] (0,0) rectangle (1.ex,1.ex);}~P9,
    \protect\raisebox{2pt}{\protect\tikz \protect\draw[gray,fill=gray] (0,0) rectangle (1.ex,1.ex);}~R5,
    \protect\raisebox{2pt}{\protect\tikz \protect\draw[gray,fill=ts5] (0,0) rectangle (1.ex,1.ex);}~W5,
    \protect\raisebox{2pt}{\protect\tikz \protect\draw[gray,fill=ts6] (0,0) rectangle (1.ex,1.ex);}~R9,
    \protect\raisebox{2pt}{\protect\tikz \protect\draw[gray,fill=ts7] (0,0) rectangle (1.ex,1.ex);}~W9,
    }
    \label{fig:masses}
    % \vspace{-10pt}
\end{figure*}

\subsection{Sensor data and annotation}

The dataset was acquired with 4 multi-sensor devices, Intel RealSense D435i, and an 8-element circular microphone array. Each D435i device has 3 cameras and provides spatially aligned RGB, narrow-baseline stereo infrared, and depth images at 30~Hz with 1280x720 pixels resolution. One D435i is mounted on a robot arm that does not move during the acquisition and provides a more realistic view of the operating area from the robot perspective. Another D435i is chest mount by the person to provide a first-person view, while the remaining two devices are placed at the sides of the robot arm as third-person views that look at the operating area. The microphone array is placed on a table and consists of 8 Boya BY-M1 omnidirectional Lavelier microphones arranged in a circular shape of radius 15~cm. Audio signals are sampled synchronously at 44.1~kHz with a multi-channel audio recorder. All signals are software-synchronized with a rate of 30~Hz. The calibration information (intrinsic and extrinsic parameters) for each  D435i and the inertial measurements of the D435i used as a body-worn camera are also provided.

The annotation of the data includes the capacity of the container, the content type, the content level, the mass of the container, the mass of the content, the maximum width and height (and depth for boxes) of each object. Fig.~\ref{fig:masses} shows the total object mass across containers and their contents.

The dataset is split into training set (684 recordings of 9 containers), public test set (228 recordings of 3 containers), and private test set (228 recordings of 3 containers). The containers for each set are evenly distributed among the three categories. The annotations of the container capacity, content type and level, and the masses of the container and content are provided publicly only for the training set.

\subsection{Tasks and performance scores}

We define three tasks for the framework, namely the classification of the amount of content (Task 1), the classification of the content type (Task 2), and the estimation of the capacity of the container (Task 3). We refer to the amount of content as {\em filling level} and to the type of content as {\em filling type}.

In Task 1, a container is either empty or filled with an unknown content at 50\% or 90\% of its capacity. There are three classes: \textit{empty}, \textit{half-full}, \textit{full}. For each configuration $j$, the goal is to classify the filling level ($\lambda^j$). 
In Task 2, containers are either empty or filled with an unknown content. There are four filling type classes:  \textit{none}, \textit{pasta}, \textit{rice}, \textit{water}.  For each configuration $j$, the goal is to classify the type of filling, if any ($\tau^j$). For these two tasks, we compute precision, recall, and F1-score for each class $k$ across all the configurations belonging to class $k$, $J_k$. \textit{Precision} is the number of true positives over the total number of true positives and false positives for each class $k$ ($P_k$). \textit{Recall} is the number of true positives over the total number of true positives and false negatives for each class $k$ ($R_k$). \textit{F1-score} is the harmonic mean of precision and recall for each class $k$ and defined as
\begin{equation}
    F_k = 2\frac{P_k R_k}{P_k + R_k}.
\end{equation}
We then compute the weighted average F1-score, $\bar{F}_1$, across the $K$ classes,
\begin{equation}
    \bar{F}_1 =\sum_{k=1}^K  \frac{J_k F_k}{J},
    \label{eq:wafs}
\end{equation}
where $J=\sum_{k=1}^K J_k$ is the total number of configuration. Note that $K=3$ for filling level classification, whereas $K=4$ for filling type classification.

In Task 3, containers vary in shape and size. For each configuration $j$, the goal is to estimate the capacity of the container ($\gamma^j \in \mathbb{R}_{>0}$, in milliliters). For capacity estimation, we compute the relative absolute error between the estimated capacity, $\tilde{\gamma}^j$, and the annotated capacity, $\gamma^j$, for each configuration, $j$, 
\begin{equation}
    \varepsilon^j = \frac{|\tilde{\gamma}^j - \gamma^j |}{\gamma^j}.
    \label{eq:ware}
\end{equation}
We then compute the average capacity score, $\bar{C}$, as
\begin{equation}
    \bar{C} = \frac{1}{J} \sum_{j=1}^J \mathds{1} e^{-\varepsilon^j},
    \label{eq:acs}
\end{equation}
where the value of the indicator function $\mathds{1} \in \{0,1\}$ is 0 only when the capacity (mass) of the container in configuration $j$ is not estimated.

The weight of the object, $\omega \in \mathbb{R}_{>0}$ (in Newtons), is the sum of the mass of the (empty) container, $m_c \in \mathbb{R}_{>0}$ (in grams), and the mass of the (unknown) filling, $m_f \in \mathbb{R}_{>0}$ (in grams), multiplied by the gravitational earth acceleration, $g=9.81$~m/s$^{-2}$,
\begin{equation}
    \omega = (m_c + m_f) g.
    \label{eq:objweight}
\end{equation}
While we do not require the mass of the empty container to be estimated, we expect methods to estimate the capacity of the container and to determine the type and amount of filling to estimate the mass of the filling. For each configuration $j$, we then compute the filling mass as 
\begin{equation}
    m_f^j = \lambda^j \gamma^j D(\tau^j),
    \label{eq:fillingmass}
\end{equation}
where $D(\cdot)$ selects a pre-computed density based on the classified filling type. The density of pasta and rice is computed from the annotation of the filling mass, capacity of the container, and filling level for each container. Density of water is 1 g/mL. For filling mass estimation, we compute the relative absolute error between the estimated, $\tilde{m}_{f}^j$, and the annotated filling mass, $m_{f}^j$, for each configuration, $j$, unless the annotated mass is zero (empty filling level),
\begin{equation}
    \epsilon^j = 
    \begin{cases}
    0, & \text{if } m_{f}^j = 0 \land \tilde{m}_{f}^j=0, \\
    \tilde{m}_{f}^j, & \text{if } m_{f}^j = 0 \land \tilde{m}_{f}^j \neq 0, \\
    \frac{|\tilde{m}_{f}^j - m_{f}^j |}{m_{f}^j}, & \text{otherwise}.
    \end{cases}
    \label{eq:ware2}
\end{equation}
Similarly to the average capacity score, we compute the average filling mass score, $\bar{M}$.

Note that we will present the scores as percentages when discussing the results in the comparative analysis. 

\subsection{Baselines}

CORSMAL provides along with the framework 12 audio-only baselines and one video-only baseline for the tasks of filling level and filling type classification.

The audio-only baselines\footnote{Baselines for audio-based classification of the content properties are available at:  \url{https://github.com/CORSMAL/CCM_ML_baselines}} jointly classify filling type and level using traditional acoustic features, such as ZCR, MFCCs, tonnetz, or spectrograms, combined with either of three machine learning classifiers (kNN, SVM, RF). Note that for MFCCs, the 1$^\text{st}$ to 13$^\text{th}$ coefficients are used, whereas the 0$^\text{th}$ coefficient is discarded. Three baselines use as input the mean and standard deviation of the MFCCs and ZCR features across multiple audio frames~\cite{Piczak2015EnvironmentalSC}. Three other baselines extract a feature vector consisting of 193 coefficients from the mean and standard deviation of the MFCCs, chromogram, Mel-scaled spectrogram, spectral contrast, and tonnetz across multiple audio frames~\cite{Vivek2020ICCSP,Shao2008RobustSI,Ghosal2018MusicGR,R2020AudioVS}. For simplicity, we refer to this set of acoustic features as AF193 in the rest of the paper. Three other baselines use spectrograms, which are cropped, resized and reshaped into a vector of dimension 9,216, as input to the classifiers~\cite{Donaher2021EUSIPCO_ACC}. To remove redundant information, three additional baselines perform dimensionality reduction with Principal Component Analysis (PCA) on the reshaped spectrograms, retaining only the first 128 components.

The vision-only baseline uses two CNNs to perform an independent classification of filling level and filling type from a single RGB image. We re-trained ResNet-18 architectures~\cite{He2016CVPR_ResNet} using a subset of frames\footnote{Data available at: \url{https://corsmal.eecs.qmul.ac.uk/filling.html}} selected within the video recordings of the training set of the CORSMAL Containers Manipulation and cropped to a rectangular area around the container~\cite{Modas2021ArXiv}. On the test sets, the baseline is applied to each camera view independently: an image crop is extracted from the last frame using Mask R-CNN~\cite{He2017ICCV_MaskRCNN} and the segmentation mask with the most confident class between \textit{cup} and \textit{wine glass} is selected. The output classes of the two CNNs include an additional class, \textit{opaque}, to handle cases where containers are not transparent and vision alone fails to determine the content type and level~\cite{Modas2021ArXiv,Mottaghi2017ICCV}.

\section{Filling level and type classification}
\label{sec:fillingclass}

Six methods used the framework to address the tasks of filling level classification (Task 1) and filling type  classification (Task 2) either independently, e.g., when only one of the two properties is necessary for the target application, or jointly, e.g., when both properties are necessary for accurately estimating the total object weight. For simplicity, we refer to the 6 methods as M1, M2~\cite{Christmann2020NTNU}, M3~\cite{Liu2020ICPR}, M4~\cite{Ishikawa2020ICPR}, M5~\cite{Iashin2020ICPR} and M6~\cite{Donaher2021EUSIPCO_ACC} for the rest of the paper. 

For filling type classification, audio is preferred as input modality and methods used either only CNNs, CNN with RNN, or CNN followed by majority voting as classification approaches~\cite{Christmann2020NTNU,Ishikawa2020ICPR,Iashin2020ICPR}. For filling level classification, some methods used visual data in combination with audio data~\cite{Iashin2020ICPR,Liu2020ICPR}. Hand-crafted and/or learned acoustic features are used by the methods.  Traditional acoustic features, such as MFCCs, spectral characteristics, ZCR, chroma vector and deviation, are computed from short-term windows. Long-term features can be obtained by summarizing the short-term features from longer windows of the input audio signal and by including additional statistics, such as mean and standard deviation. Learned features are extracted by CNNs from multi-channel or mono-channel audio signals that are post-processed into spectrograms or log-Mel spectrograms~\cite{Ishikawa2020ICPR,Iashin2020ICPR}. To handle audio signals of different duration, long audio signals can be truncated to a pre-defined duration and zero-padding is added to shorter signals~\cite{Christmann2020NTNU,Ishikawa2020ICPR}.

The fully connected neural network of M1 has 5 layers and uses STFT features as input. The network is trained with the Adam optimizer~\cite{Kingma2015ICLR_ADAM} and dropout~\cite{Srivastava2014DropoutAS} on the last hidden layer to reduce overfitting. 

The filling type classifier of M2 uses 40 normalized and concatenated MFCCs features that are extracted with 20~ms windows at 22~kHz, with a maximum duration of 30~s~\cite{Christmann2020NTNU}.
The CNN has 2 convolutional layers and 1 fully connected layer (86,876 trainable parameters). 

M4~\cite{Ishikawa2020ICPR} used all the 8 audio signals from the microphone array to compute log Mel-scaled spectrograms with STFT and 64 filter banks for filling type and filling level classification. A sliding window over the cropped spectrogram with 75\% overlap forms overlapping audio frames consisting of 3D tensors, where the third dimension is given by the 8 audio channels. Each window is provided as input to a CNN consisting of 5 blocks, each with 2 convolutional and 1 batch normalization layers followed by a max-pooling layer. The CNN is complemented by 3 fully connected layers for the filling type classification of each audio frame and followed by the majority voting. The CNN has a total of 13 layers with 4,472,580 trainable parameters. The same extracted features are also used as input to the three stacked Long Short-Term Memory (LSTM)~\cite{Hochreiter1997LSTM} units for the filling level classification. The three stacked LSTMs are trained with a set of 100 audio frames and contain 256 hidden states, resulting in 2,366,211 trainable parameters.

The multi-layer perceptrons (MLPs) of M3~\cite{Liu2020ICPR} are trained for either filling level or filling type classification, and specifically only for each object category (\textit{cup}, \textit{drinking glass}, \textit{food box}). Each MLP has 3 layers with 3,096 nodes in the first hidden layer and 512 in the last hidden layer. The total number of trainable parameters is 20,762,288. The MLPs takes as input a spectrogram computed from a multi-channel sound signal re-sampled at 16,600~Hz and converted into mono-channel by averaging the samples across channels. Only the last 32,000 samples are retained and converted into a spectrogram via Discrete Fourier Transform.
To select which MLP to use at inference time, regions of interest (ROIs) are detected in all frames of the image sequences of all four views in the CORSMAL Containers Manipulation dataset by using YOLOv4~\cite{Bochkovskiy2020_YOLOv4} pre-trained on MS COCO~\cite{Lin2018ECCV_COCO}. The category (\textit{cup}, \textit{drinking glass}, \textit{food box}) is determined by a majority voting of randomly sampled frames (65\% of all frames). 

Both traditional and learned acoustic features are used by M5~\cite{Iashin2020ICPR} for filling type classification, whereas visual features are extracted in addition to the acoustic features for filling level classification. Multiple classifiers, each associated with each feature, are used to output the class probabilities. Then, the probabilities are averaged across the classifiers to determine the final class. For the acoustic features, the multi-channel input audio signal is converted into a mono-channel by averaging the samples across channels. MFCCs, energy, spectral characteristics, and their statistics (mean and standard deviation) are computed from 50~ms windows of the input signal as short-term traditional features. The features are concatenated in a 136-dimensional vector used as input to a RF classifier. The number of trees of the RF classifier is automatically set during training by selecting the value between $(10, 25, 50, 100, 200, 500)$ that achieves the highest accuracy in validation. For the learned features, the mono-channel signal is re-sampled  at 16~kHz and converted into log-Mel spectrograms from 960~ms windows of the re-sampled signal. Each spectrogram is provided as input to a VGG-based model \cite{hershey2017cnn} that is pre-trained on a large dataset (e.g.,  AudioSet~\cite{45857}) and computes a 128-dimensional feature vector. The learned features are then provided as input to a GRU model~\cite{chung2014empirical} that has 5 layers and a hidden layer of size 512 to handle the intrinsic temporal relations of the signals. The model has a total of 7,291,395 trainable parameters. 
Visual features are extracted from the image sequences of all camera views by using R(2+1)D~\cite{Tran2018CVPR_R21D}, a spatio-temporal CNN that is based on residual connections~\cite{He2016CVPR_ResNet} and 18 (2+1)D convolutional layers that approximate 3D convolution by a 2D convolution (spatial) followed by a 1D convolution (temporal). R(2+1)D is pre-trained for action recognition on Kinetics 400~\cite{kay2017kinetics}, takes as input a fixed window of 16 RGB frames of 112$\times$112 pixel resolution, and outputs a 512-dimensional feature vector. Long temporal relations between the features of each window are estimated by using a RNN with a GRU model  that has 3 layers and a hidden dimension of size 512 (4,729,347 trainable parameters). The GRU models from each camera view are  jointly trained and their logits are  summed together before applying the final softmax to obtain the class probabilities from the visual input. For filling type classification, the probabilities resulting from the last hidden state of the GRU network and those resulting from the RF are averaged. For filling level classification, the probabilities resulting from the RF classifier and the GRU models for both the audio and visual features are averaged together to compute the final class. The RF classifier and all the GRU models are trained independently for filling type classification and filling level classification by using 3-fold validation strategy.

Jointly estimating the filling type and level can avoid infeasible cases, such as an \textit{empty}  \textit{water} or \textit{half-full}  \textit{none}. Different traditional classifiers and existing CNNs that use spectrograms as input have been analyzed and compared in Donaher et al.'s work~\cite{Donaher2021EUSIPCO_ACC}, especially when different containers are manipulated by a person with different content types, such as both liquids and granular materials.

Because of the different container types and corresponding manipulation, the authors of M6~\cite{Donaher2021EUSIPCO_ACC} decomposed the problem into two steps, namely action recognition and content classification and devised three independent CNNs. The first CNN (action classifier) identifies the manipulation performed by the human, i.e., shaking or pouring, and the other two CNNs are task-specific and determine the filling type and level. The CNN for action recognition (\textit{pouring}, \textit{shaking}, \textit{unknown}) has 4 convolutional, 2 max-pooling, and 3 fully connected layers; the CNN for the specific action of pouring has 6 convolutional, 3 max-pooling, and 3 fully connected layers; and the CNN for the specific action of shaking has 4 convolutional, 2 max-pooling, and 2 fully connected layers. The choice of which task-specific network should be used is conditioned by the decision of the first CNN. When the action classifier does not distinguish between pouring or shaking, the approach associates the \textit{unknown} case to the class \textit{empty}.

\section{Capacity estimation}
\label{sec:capacity}

We categorize the methods as regression~\cite{Christmann2020NTNU,Liu2020ICPR} and geometric-based approaches~\cite{Ishikawa2020ICPR,Iashin2020ICPR}. These methods use either RGB, RGB and depth data, or multiple RGB images from the CORSMAL Containers Manipulation dataset.  

Regression approaches use CNNs~\cite{Christmann2020NTNU} or distribution fitting via Gaussian processes~\cite{Liu2020ICPR}. The CNN architecture of M2 has 4 convolutional layers, each followed by batch normalization~\cite{Ioffe2015ICML_BatchNorm}, and 3 fully connected layers (532,175 trainable parameters)~\cite{Christmann2020NTNU}. The CNN takes as input a ROI and its normalized relative size, and then regresses the capacity of the container limited to 4,000 mL, accordingly to the range of capacities in the dataset. The ROI is computed from the contour features of a depth image selected from the frame with the most visible pixels of the frontal, fixed view and assuming a maximum depth of 700~mm. 
M4~\cite{Liu2020ICPR} used Gaussian processes to regress the container capacity, depending on the container category. To model multiple multi-variate Gaussian functions for each container category, the container type is recognized by detecting multiple ROIs in all frames of all image sequences as done for filling type and level classification.

Geometric-based approaches approximate the container to a primitive shape in 3D, such as cuboid or cylinder\cite{Ishikawa2020ICPR,Xompero2020ICASSP_LoDE,Iashin2020ICPR}. The shape is represented as a point cloud obtained directly from RGB-D data or computed via energy-based minimization to fit the points to the real shape of the object as observed in the RGB images of a wide-baseline stereo camera and constrained by the object masks~\cite{Xompero2020ICASSP_LoDE,Iashin2020ICPR}. The capacity is then computed as a by-product, e.g., by finding the minimum and maximum values for each coordinate in 3D~\cite{Ishikawa2020ICPR} or using volume formulas specific for the primitive shape~\cite{Iashin2020ICPR}. 
The approximated primitives can lead to inaccurate capacities: a cuboid representation could result in an overestimated capacity and hence re-scaling would be necessary~\cite{Ishikawa2020ICPR}; a cylinder representation may not generalize to different shapes than rotationally symmetric objects. 
To handle occlusions caused by the human hand manipulating a container, M5~\cite{Ishikawa2020ICPR} selects the RGB-D frame with a single silhouette having the largest number of pixels and post-processes the point cloud to deal with inaccuracies in the segmentation. Capacity estimations computed at different frames of the image sequences in the stereo views are then averaged, assuming that the container is fully visible.

%%%%%%%%%%%%%%%%%%%%%%%%%%%%%%%%%
%%%%%%%%%%%%%%%%%%%%%%%%%%%%%%
\section{Experiment results and discussion}
\label{sec:analysis}

We compare and analyze the performance of the 6 methods and the 13 baselines on the public test set, the private test set, and their combination on the CORSMAL Containers Manipulation dataset~\cite{Xompero_CCM}. 

\begin{figure*}[t!]
    \centering
    \includegraphics[width=0.8\linewidth]{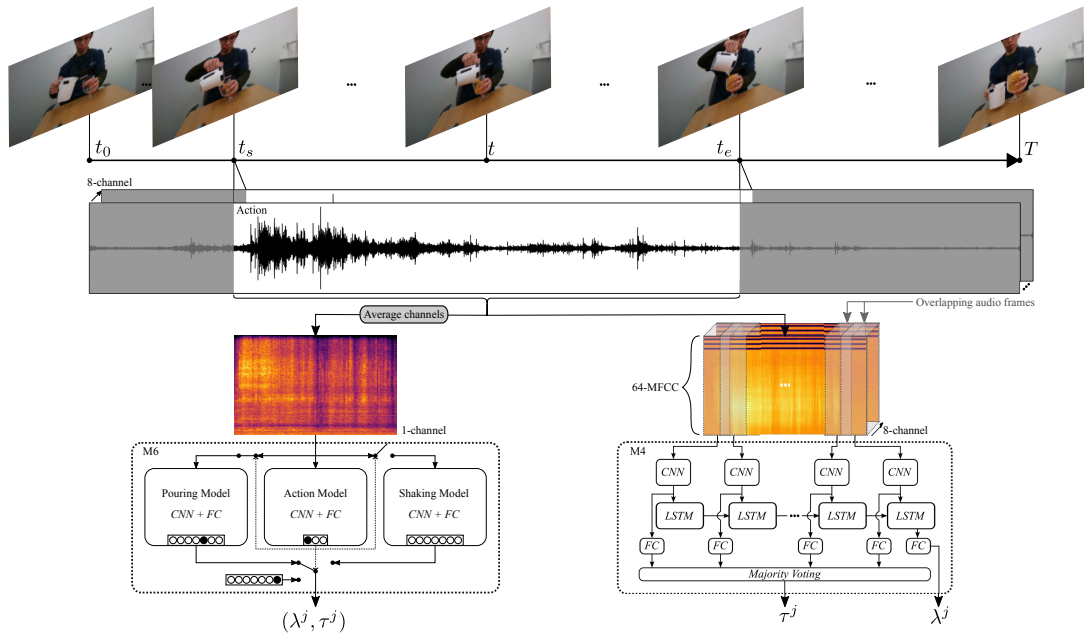}
    \caption{Illustrative comparison of M6~\cite{Donaher2021EUSIPCO_ACC} (left) and M4~\cite{Ishikawa2020ICPR} (right) for filling type ($\tau^j$) and level classification ($\lambda^j$). The two methods take as input only an audio signal that is converted into a spectrogram representation. During training, the initial and final part of the audio signal (gray areas) are removed based on the manual annotations and to focus only on the action. Note that M4~\cite{Ishikawa2020ICPR} (right) computes MFCC features from overlapping audio frames (shadow gray areas on the spectrogram). KEY -- CNN:~convolutional neural network, FC:~fully connected layer, LSTM:~Long-Short Term Memory, MFCC:~Mel Frequency Cepstral Coefficients.
    }
    \label{fig:accvshvrl}
    % \vspace{-10pt}
\end{figure*}

\subsection{Implementation details}

The CNN of M2 for filling type classification is trained with the SGD optimizer, a fixed learning rate of 0.00025 and momentum of 0.9, and a batch size of 16. M4 sets the frame length to 25~ms, the hop-length to 10~ms, and the number of samples for the Fast Fourier Transform to 512 for computing the STFT. During training, M4 crops audio signals based on manual annotations of the starting and ending of the manipulation. The network for filling level classification of M4 is trained by using cross-entropy loss and the Adam optimizer~\cite{Kingma2015ICLR_ADAM} with a learning rate of 0.00001 and a mini-batch size of 32 for 200 epochs.

\subsection{Filling level classification}

Table~\ref{tab:leaderboardtask1} compares the performance of all baselines and methods except M2. M4, M5 and M6 achieve the highest accuracy with 80.84, 79.65, and 78.65~$\bar{F}_1$ on the combined test set, respectively. This performance is almost twice higher than M1 and M3 and shows that using only audio as input modality is sufficient to achieve an accuracy higher than 75~$\bar{F}_1$. M5 uses both audio and visual data, but the similar performance to M4 and M6 suggests that audio features are dominant in determining the classification decision. 
M6 is the best performing in the private test set (81.46~$\bar{F}_1$), whereas M4 is the best performing in the public test set (82.63~$\bar{F}_1$). Interestingly, both methods selected a fixed portion of the audio signal, transformed into a spectrogram, where the manipulation of the container by the human subject was more likely to occur (see Fig.~\ref{fig:accvshvrl}). However, the three CNNs of M6 use the full trimmed spectrograms as input, whereas the CNN+LSTM of M4 uses portions of the log-Mel spectrogram, which are obtained with a temporal sliding window. Both are shown to be valid methods assuming that the whole audio signal is available and the manipulation is completed.

\begin{table}[b!]
    \centering
    \setlength\tabcolsep{1.5pt}
    \caption{Filling level classification results (Task 1). Baselines and state-of-the-art methods (MX with X ranges from 1 to 6) are ranked by their score in the combined test set.  
    }
    \begin{tabular}{lcccccccc}
    \specialrule{1.2pt}{0.2pt}{1pt}
    \textbf{Method} & \multicolumn{5}{c}{\textbf{Input modality}} & \multicolumn{3}{c}{\textbf{Test set}}  \\
    \cmidrule(lr){2-6}\cmidrule(lr){7-9}
     & A & R1 & R2 & R3 & R4 & Public & Private & Combined  \\
    \specialrule{1.2pt}{0.2pt}{1pt}
    Mask + RN & \wbox & \wbox & \wbox & \wbox & \bbox & 25.12 & 21.99 & 23.68 \\
    Mask + RN & \wbox & \wbox & \wbox & \bbox & \wbox & 36.52 & 25.52 & 31.46 \\
    Spect. + PCA + SVM  & \bbox & \wbox & \wbox & \wbox & \wbox & 30.08 & 31.99 & 31.64 \\
    \rowcolor{mylightgray}
    Random & -- & -- & -- & -- & -- & 33.35 & 41.86 & 37.62 \\
    Spect. + PCA + kNN  & \bbox & \wbox & \wbox & \wbox & \wbox & 39.03 & 37.16 & 38.31 \\
    Mask + RN & \wbox & \wbox & \bbox & \wbox & \wbox & 48.90 & 26.73 & 39.00 \\
    M3~\cite{Liu2020ICPR}         & \bbox & \bbox & \bbox & \bbox & \bbox & 44.31 & 42.70 & 43.53 \\
    Spect. + PCA + RF   & \bbox & \wbox & \wbox & \wbox & \wbox & 46.79 & 42.46 & 44.66 \\
    Spect. + RF       & \bbox & \wbox & \wbox & \wbox & \wbox & 45.43 & 45.59 & 45.49 \\
    Mask + RN & \wbox & \bbox & \wbox & \wbox & \wbox & 58.51 & 32.93 & 47.00 \\
    M1     & \bbox & \wbox & \wbox & \wbox & \wbox & 50.73 & 47.08 & 48.71 \\
    Spect. + SVM      & \bbox & \wbox & \wbox & \wbox & \wbox & 47.66 & 51.54 & 49.67 \\
    AF193 + kNN         & \bbox & \wbox & \wbox & \wbox & \wbox & 55.49 & 53.22 & 54.47 \\
    Spect. + kNN      & \bbox & \wbox & \wbox & \wbox & \wbox & 59.15 & 53.47 & 56.38 \\
    ZCR + MFCCs + kNN    & \bbox & \wbox & \wbox & \wbox & \wbox & 63.63 & 54.97 & 59.35 \\
    AF193 + SVM         & \bbox & \wbox & \wbox & \wbox & \wbox & 60.77 & 58.57 & 60.09 \\
    ZCR + MFCCs + SVM    & \bbox & \wbox & \wbox & \wbox & \wbox & 66.27 & 57.19 & 61.87 \\
    AF193 + RF          & \bbox & \wbox & \wbox & \wbox & \wbox & 64.18 & 63.94 & 64.74 \\
    ZCR + MFCCs + RF     & \bbox & \wbox & \wbox & \wbox & \wbox & 70.04 & 63.11 & 66.80 \\
    M4~\cite{Ishikawa2020ICPR}            & \bbox & \wbox & \wbox & \wbox & \wbox & \textbf{82.63} & 74.43 & 78.56 \\
    M5~\cite{Iashin2020ICPR} & \bbox & \bbox & \bbox & \bbox & \bbox & 78.14 & 81.16 & 79.65 \\
    M6~\cite{Donaher2021EUSIPCO_ACC} & \bbox & \wbox & \wbox & \wbox & \wbox & 80.22 & \textbf{81.46} & \textbf{80.84} \\
    \specialrule{1.2pt}{0.2pt}{1pt}
    \multicolumn{9}{l}{\scriptsize{Bold font: result of the best performing method.}}\\
    \multicolumn{9}{l}{\scriptsize{\parbox{0.85\columnwidth}{KEY -- A:~audio, RX:~RGB for view X (1,2,3,4), Mask + RN:~Mask R-CNN + ResNet-18, ZCR:~zero crossing rate, MFCCs:~Mel-frequency cepstrum coefficients, Spect.:~spectrogram, RF:~random forest, SVM:~support vector machine, kNN:~k-nearest neighbor, PCA:~principal component analysis, AF193:~193 audio features consisting of MFCCs, chromogram, Mel-scaled spectrogram, spectral contrast, and tonal centroid.}}} \\
    \end{tabular}
    \label{tab:leaderboardtask1}
    % \vspace{-10pt}
\end{table}

%%%%%%%%%%%%%%%%%%%%%%%%%%%%%%%%%%%
% Public test Data
\pgfplotstableread{confmat.txt}\confmat

\pgfplotstableread{confmat_combined_bit_level.txt}\confmatbitlevel
\pgfplotstableread{confmat_combined_hvrl_level.txt}\confmathvrllevel
\pgfplotstableread{confmat_combined_concatenation_level.txt}\confmatconcatlevel
\pgfplotstableread{confmat_combined_challengers_level.txt}\confmatchallengerslevel
\pgfplotstableread{confmat_combined_mrn1_level.txt}\confmatmrnlevel
\pgfplotstableread{confmat_combined_acc_level.txt}\confmatacclevel

\begin{figure*}[t!]
    \centering
    %%%% VA2MASS
    \begin{tikzpicture}
        \begin{axis}[
            enlargelimits=false,
            axis on top,
            axis x line*=bottom,
            width=.20\linewidth,
            ylabel=True label,
            xlabel=\textbf{Mask + RN},
            view={0}{90},
            y dir=reverse,
            point meta min=0,
            point meta max=1,
            xmin=-0.5,xmax=2.5,
            ymin=-0.5,ymax=2.5,
            ytick={0,1,2},
            xtick={0,1,2},
            xticklabels={},
            yticklabels={E,H,F},
            tick label style={font=\scriptsize},
            label style={font=\footnotesize},
            colormap={bw}{gray(0cm)=(1);gray(1cm)=(0);},
        ]
        \addplot [matrix plot*,mesh/cols=3, point meta=explicit] table [meta=z] {\confmatmrnlevel};
        \end{axis}
        \begin{axis}[
            axis x line*=top,
            axis y line*=right,
            y dir=reverse,
            width=.20\linewidth,
            view={0}{90},   % not needed for `matrix plot*' variant
            xlabel=Predicted label,
            enlargelimits=false,
            axis on top,
            point meta min=0,
            point meta max=1,
            xmin=-0.5,xmax=2.5,
            ymin=-0.5,ymax=2.5,
            ytick={0,1,2},
            xtick={0,1,2},
            yticklabels={},
            xticklabels={E,H,F},
            tick label style={font=\scriptsize},
            label style={font=\footnotesize},
         ]
         \end{axis}
    \end{tikzpicture}
    %%%% Challengers
   \begin{tikzpicture}
        \begin{axis}[
            enlargelimits=false,
            axis on top,
            axis x line*=bottom,
            width=.20\linewidth,
            % ylabel=True label,
            xlabel=\textbf{M1},
            view={0}{90},
            y dir=reverse,
            point meta min=0,
            point meta max=1,
            xmin=-0.5,xmax=2.5,
            ymin=-0.5,ymax=2.5,
            ytick={0,1,2},
            xtick={0,1,2},
            xticklabels={},
            yticklabels={E,H,F},
            tick label style={font=\scriptsize},
            label style={font=\footnotesize},
            colormap={bw}{gray(0cm)=(1);gray(1cm)=(0);},
        ]
        \addplot [matrix plot*,mesh/cols=3, point meta=explicit] table [meta=z] {\confmatchallengerslevel};
        \end{axis}
        \begin{axis}[
            axis x line*=top,
            axis y line*=right,
            y dir=reverse,
            width=.20\linewidth,
            view={0}{90},   % not needed for `matrix plot*' variant
            xlabel=Predicted label,
            enlargelimits=false,
            axis on top,
            point meta min=0,
            point meta max=1,
            xmin=-0.5,xmax=2.5,
            ymin=-0.5,ymax=2.5,
            ytick={0,1,2},
            xtick={0,1,2},
            yticklabels={},
            xticklabels={E,H,F},
            tick label style={font=\scriptsize},
            label style={font=\footnotesize},
         ]
         \end{axis}
    \end{tikzpicture}
    %%%% VA2MASS
    \begin{tikzpicture}
        \begin{axis}[
            enlargelimits=false,
            axis on top,
            axis x line*=bottom,
            width=.20\linewidth,
            % ylabel=True label,
            xlabel=\textbf{M3},
            view={0}{90},
            y dir=reverse,
            point meta min=0,
            point meta max=1,
            xmin=-0.5,xmax=2.5,
            ymin=-0.5,ymax=2.5,
            ytick={0,1,2},
            xtick={0,1,2},
            xticklabels={},
            yticklabels={E,H,F},
            tick label style={font=\scriptsize},
            label style={font=\footnotesize},
            colormap={bw}{gray(0cm)=(1);gray(1cm)=(0);},
        ]
        \addplot [matrix plot*,mesh/cols=3, point meta=explicit] table [meta=z] {\confmatconcatlevel};
        \end{axis}
        \begin{axis}[
            axis x line*=top,
            axis y line*=right,
            y dir=reverse,
            width=.20\linewidth,
            view={0}{90},   % not needed for `matrix plot*' variant
            xlabel=Predicted label,
            enlargelimits=false,
            axis on top,
            point meta min=0,
            point meta max=1,
            xmin=-0.5,xmax=2.5,
            ymin=-0.5,ymax=2.5,
            ytick={0,1,2},
            xtick={0,1,2},
            yticklabels={},
            xticklabels={E,H,F},
            tick label style={font=\scriptsize},
            label style={font=\footnotesize},
         ]
         \end{axis}
    \end{tikzpicture}
    %%%% HVRL
    \begin{tikzpicture}
          \begin{axis}[
            enlargelimits=false,
            axis on top,
            axis x line*=bottom,
            width=.20\linewidth,
            % ylabel=True label,
            xlabel=\textbf{M4},
            view={0}{90},
            y dir=reverse,
            point meta min=0,
            point meta max=1,
            xmin=-0.5,xmax=2.5,
            ymin=-0.5,ymax=2.5,
            ytick={0,1,2},
            xtick={0,1,2},
            xticklabels={},
           yticklabels={E,H,F},
            tick label style={font=\scriptsize},
            label style={font=\footnotesize},
            colormap={bw}{gray(0cm)=(1);gray(1cm)=(0);},
        ]
        \addplot [matrix plot*,mesh/cols=3, point meta=explicit] table [meta=z] {\confmathvrllevel};
        \end{axis}
        \begin{axis}[
            axis x line*=top,
            axis y line*=right,
            y dir=reverse,
            width=.20\linewidth,
            view={0}{90},   % not needed for `matrix plot*' variant
            xlabel=Predicted label,
            enlargelimits=false,
            axis on top,
            point meta min=0,
            point meta max=1,
            xmin=-0.5,xmax=2.5,
            ymin=-0.5,ymax=2.5,
            ytick={0,1,2},
            xtick={0,1,2},
            yticklabels={},
            xticklabels={E,H,F},
            tick label style={font=\scriptsize},
            label style={font=\footnotesize},
         ]
         \end{axis}
    \end{tikzpicture}
    %%%% BIT
    \begin{tikzpicture}
          \begin{axis}[
            enlargelimits=false,
            axis on top,
            axis x line*=bottom,
            width=.20\linewidth,
            % ylabel=True label,
            xlabel=\textbf{M5},
            view={0}{90},
            y dir=reverse,
            point meta min=0,
            point meta max=1,
            xmin=-0.5,xmax=2.5,
            ymin=-0.5,ymax=2.5,
            ytick={0,1,2},
            xtick={0,1,2},
            xticklabels={},
            yticklabels={E,H,F},
            tick label style={font=\scriptsize},
            label style={font=\footnotesize},
            colormap={bw}{gray(0cm)=(1);gray(1cm)=(0);},
        ]
        \addplot [matrix plot*,mesh/cols=3, point meta=explicit] table [meta=z] {\confmatbitlevel};
        \end{axis}
        \begin{axis}[
            axis x line*=top,
            axis y line*=right,
            y dir=reverse,
            width=.20\linewidth,
            view={0}{90},   % not needed for `matrix plot*' variant
            xlabel=Predicted label,
            enlargelimits=false,
            axis on top,
            point meta min=0,
            point meta max=1,
            xmin=-0.5,xmax=2.5,
            ymin=-0.5,ymax=2.5,
            ytick={0,1,2},
            xtick={0,1,2},
            yticklabels={},
            xticklabels={E,H,F},
            tick label style={font=\scriptsize},
            label style={font=\footnotesize},
         ]
         \end{axis}
    \end{tikzpicture}
    %%%% ACC
   \begin{tikzpicture}
        \begin{axis}[
            enlargelimits=false,
            axis on top,
            axis x line*=bottom,
            width=.20\linewidth,
            % ylabel=True label,
            xlabel=\textbf{M6},
            view={0}{90},
            y dir=reverse,
            point meta min=0,
            point meta max=1,
            xmin=-0.5,xmax=2.5,
            ymin=-0.5,ymax=2.5,
            ytick={0,1,2},
            xtick={0,1,2},
            xticklabels={},
            yticklabels={E,H,F},
            tick label style={font=\scriptsize},
            label style={font=\footnotesize},
            % colormap={bw}{gray(0cm)=(1);gray(1cm)=(0);},
            colorbar,
            colorbar style={
                width=.02\columnwidth, 
                % at={(3,1.05)},
                % anchor=west,
                % yticklabel pos=lower,
                yticklabel style={
                    /pgf/number format/.cd,
                    fixed,
                    precision=1,
                    fixed zerofill,
                    tick label style={font=\scriptsize},
                },
            },
            colormap={bw}{gray(0cm)=(1);gray(1cm)=(0);},
        ]
        \addplot [matrix plot*,mesh/cols=3, point meta=explicit] table [meta=z] {\confmatacclevel};
        \end{axis}
        \begin{axis}[
            axis x line*=top,
            axis y line*=right,
            y dir=reverse,
            width=.20\linewidth,
            view={0}{90},   % not needed for `matrix plot*' variant
            xlabel=Predicted label,
            enlargelimits=false,
            axis on top,
            point meta min=0,
            point meta max=1,
            xmin=-0.5,xmax=2.5,
            ymin=-0.5,ymax=2.5,
            ytick={0,1,2},
            xtick={0,1,2},
            yticklabels={},
            xticklabels={E,H,F},
            tick label style={font=\scriptsize},
            label style={font=\footnotesize},
         ]
         \end{axis}
    \end{tikzpicture}
    % \vspace{7pt}
    % \\
    % \begin{tikzpicture}
    % \begin{axis}[
    %         hide axis,  
    %         scale only axis,
    %         height=.05\columnwidth, width=0.5\columnwidth,
    %         % enlargelimits=false,
    %         % at={(frame66a.north west)}, 
    %         anchor=south west,
    %         point meta min=0,
    %         point meta max=1,
    %         colorbar horizontal,
    %         colorbar style={
    %             height=.02\columnwidth, 
    %             at={(3,1.05)},
    %             anchor=south west,
    %             xticklabel pos=lower,
    %             xticklabel style={
    %                 /pgf/number format/.cd,
    %                 fixed,
    %                 precision=1,
    %                 fixed zerofill,
    %                 tick label style={font=\scriptsize},
    %             },
    %             title={Counting normalised by true lables},
    %             title style={font=\footnotesize},
    %         },
    %         colormap={bw}{gray(0cm)=(1);gray(1cm)=(0);},
    %     ]
    %     \addplot [draw=none] coordinates {(0,0)};
    %     \end{axis}
    % \end{tikzpicture}
    \caption{Confusion matrices of filling level classification for all methods across all the containers of the public and private testing splits of the CORSMAL Container Manipulation dataset~\cite{Xompero_CCM}. Note that the counting for each cell is normalized by the total number of true labels for each class (colorbar). KEY --E:~empty; H:~half-full; F:~full, Mask + RN:~Mask R-CNN + ResNet-18.
    \vspace{-10pt}
    }
    \label{fig:confmatlevel}
\end{figure*}
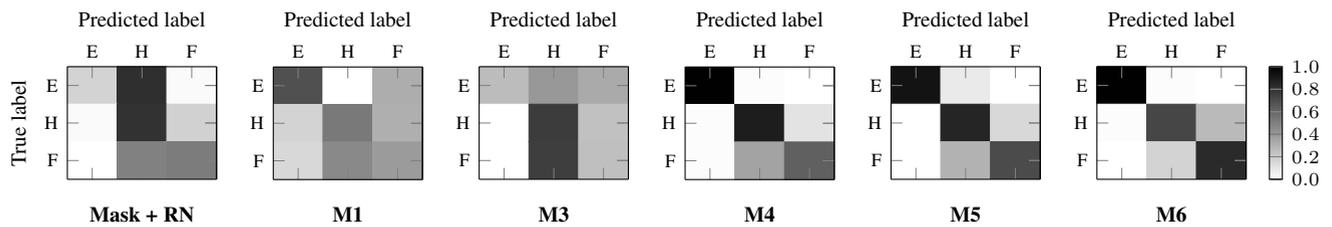

%%%%%%%%%%%%%%%%%%%%%%%%%%%%%%%%%%%

The confusion matrices in Fig.~\ref{fig:confmatlevel} show that M4 and M6 do not confuse the class \textit{empty}, whereas M5 mis-classifies some \textit{empty} configurations as \textit{half-full}. Not surprisingly, most of the confusions occur between the classes \textit{half-full} and \textit{full} for all methods. M4 and M5 are more accurate than M6 in recognizing the class \textit{half-full}, but M6 is more accurate in recognizing the class \textit{full}. M3 mis-classifies the true class \textit{empty} as \textit{half-full} for 40\% of the times and as \textit{full} for 33\% of the times, and the class \textit{full} is confused with \textit{half-full} for 75\% of the times. M3 recognizes the container categories \textit{cup}, \textit{drinking glass} and \textit{food box} with 92\%, 73\%, and 88\% accuracy, respectively, in the training set. Errors in the category recognition may lead to wrong classifications by the selected category-specific MLP-based classifier, which is also trained with limited and selected data. The CNN of M1 made erroneous predictions across all classes, except for \textit{empty} that was never predicted as \textit{half-full} but only confused with \textit{full}. The vision-only baseline (using the first camera view, on the left side of the robot arm) confused 81\% of the times the class \textit{empty} with \textit{half-full} in addition to mis-classification between \textit{half-full} and \textit{full}, making the performance of the baseline only 10~$\bar{F}_1$ points higher than a random classifier (37.62~$\bar{F}_1$). 

% Public test Data
\pgfplotstableread{confmat.txt}\confmat

\pgfplotstableread{confmat_combined_bit_type.txt}\confmatbittype
\pgfplotstableread{confmat_combined_hvrl_type.txt}\confmathvrltype
\pgfplotstableread{confmat_combined_concatenation_type.txt}\confmatconcattype
\pgfplotstableread{confmat_combined_challengers_type.txt}\confmatchallengerstype
\pgfplotstableread{confmat_combined_ntnuerc_type.txt}\confmatntnuerctype
\pgfplotstableread{confmat_combined_acc_type.txt}\confmatacctype

\begin{figure*}[t!]
    \centering
    %%%% Challengers
   \begin{tikzpicture}
        \begin{axis}[
            enlargelimits=false,
            axis on top,
            axis x line*=bottom,
            width=.19\linewidth,
            ylabel=True label,
            xlabel=\textbf{M1},
            view={0}{90},
            y dir=reverse,
            point meta min=0,
            point meta max=1,
            xmin=-0.5,xmax=3.5,
            ymin=-0.5,ymax=3.5,
            ytick={0,1,2,3},
            xtick={0,1,2,3},
            xticklabels={},
            yticklabels={E,P,R,W},
            tick label style={font=\scriptsize},
            label style={font=\footnotesize},
            colormap={bw}{gray(0cm)=(1);gray(1cm)=(0);},
        ]
        \addplot [matrix plot*,mesh/cols=4, point meta=explicit] table [meta=z] {\confmatchallengerstype};
        \end{axis}
        \begin{axis}[
            axis x line*=top,
            axis y line*=right,
            y dir=reverse,
            width=.19\linewidth,
            view={0}{90},   % not needed for `matrix plot*' variant
            xlabel=Predicted label,
            enlargelimits=false,
            axis on top,
            point meta min=0,
            point meta max=1,
            xmin=-0.5,xmax=3.5,
            ymin=-0.5,ymax=3.5,
            ytick={0,1,2,3},
            xtick={0,1,2,3},
            yticklabels={},
            xticklabels={E,P,R,W},
            tick label style={font=\scriptsize},
            label style={font=\footnotesize},
         ]
         \end{axis}
    \end{tikzpicture}
     %%%% VA2MASS
    \begin{tikzpicture}
        \begin{axis}[
            enlargelimits=false,
            axis on top,
            axis x line*=bottom,
            width=.19\linewidth,
            % ylabel=True label,
            xlabel=\textbf{M2},
            view={0}{90},
            y dir=reverse,
            point meta min=0,
            point meta max=1,
            xmin=-0.5,xmax=3.5,
            ymin=-0.5,ymax=3.5,
            ytick={0,1,2,3},
            xtick={0,1,2,3},
            xticklabels={},
            yticklabels={E,P,R,W},
            tick label style={font=\scriptsize},
            label style={font=\footnotesize},
            colormap={bw}{gray(0cm)=(1);gray(1cm)=(0);},
        ]
        \addplot [matrix plot*,mesh/cols=4, point meta=explicit] table [meta=z] {\confmatntnuerctype};
        \end{axis}
        \begin{axis}[
            axis x line*=top,
            axis y line*=right,
            y dir=reverse,
            width=.19\linewidth,
            view={0}{90},   % not needed for `matrix plot*' variant
            xlabel=Predicted label,
            enlargelimits=false,
            axis on top,
            point meta min=0,
            point meta max=1,
            xmin=-0.5,xmax=3.5,
            ymin=-0.5,ymax=3.5,
            ytick={0,1,2,3},
            xtick={0,1,2,3},
            yticklabels={},
            xticklabels={E,P,R,W},
            tick label style={font=\scriptsize},
            label style={font=\footnotesize},
         ]
         \end{axis}
    \end{tikzpicture}
        %%%% VA2MASS
    \begin{tikzpicture}
        \begin{axis}[
            enlargelimits=false,
            axis on top,
            axis x line*=bottom,
            width=.19\linewidth,
            % ylabel=True label,
            xlabel=\textbf{M3},
            view={0}{90},
            y dir=reverse,
            point meta min=0,
            point meta max=1,
            xmin=-0.5,xmax=3.5,
            ymin=-0.5,ymax=3.5,
            ytick={0,1,2,3},
            xtick={0,1,2,3},
            xticklabels={},
            yticklabels={E,P,R,W},
            tick label style={font=\scriptsize},
            label style={font=\footnotesize},
            colormap={bw}{gray(0cm)=(1);gray(1cm)=(0);},
        ]
        \addplot [matrix plot*,mesh/cols=4, point meta=explicit] table [meta=z] {\confmatconcattype};
        \end{axis}
        \begin{axis}[
            axis x line*=top,
            axis y line*=right,
            y dir=reverse,
            width=.19\linewidth,
            view={0}{90},   % not needed for `matrix plot*' variant
            xlabel=Predicted label,
            enlargelimits=false,
            axis on top,
            point meta min=0,
            point meta max=1,
            xmin=-0.5,xmax=3.5,
            ymin=-0.5,ymax=3.5,
            ytick={0,1,2,3},
            xtick={0,1,2,3},
            yticklabels={},
            xticklabels={E,P,R,W},
            tick label style={font=\scriptsize},
            label style={font=\footnotesize},
         ]
         \end{axis}
    \end{tikzpicture}
    %%%% HVRL
    \begin{tikzpicture}
          \begin{axis}[
            enlargelimits=false,
            axis on top,
            axis x line*=bottom,
            width=.19\linewidth,
            % ylabel=True label,
            xlabel=\textbf{M4},
            view={0}{90},
            y dir=reverse,
            point meta min=0,
            point meta max=1,
            xmin=-0.5,xmax=3.5,
            ymin=-0.5,ymax=3.5,
            ytick={0,1,2,3},
            xtick={0,1,2,3},
            xticklabels={},
           yticklabels={E,P,R,W},
            tick label style={font=\scriptsize},
            label style={font=\footnotesize},
            colormap={bw}{gray(0cm)=(1);gray(1cm)=(0);},
        ]
        \addplot [matrix plot*,mesh/cols=4, point meta=explicit] table [meta=z] {\confmathvrltype};
        \end{axis}
        \begin{axis}[
            axis x line*=top,
            axis y line*=right,
            y dir=reverse,
            width=.19\linewidth,
            view={0}{90},   % not needed for `matrix plot*' variant
            xlabel=Predicted label,
            enlargelimits=false,
            axis on top,
            point meta min=0,
            point meta max=1,
            xmin=-0.5,xmax=3.5,
            ymin=-0.5,ymax=3.5,
            ytick={0,1,2,3},
            xtick={0,1,2,3},
            yticklabels={},
            xticklabels={E,P,R,W},
            tick label style={font=\scriptsize},
            label style={font=\footnotesize},
         ]
         \end{axis}
    \end{tikzpicture}
        %%%% KNN
    \begin{tikzpicture}
          \begin{axis}[
            enlargelimits=false,
            axis on top,
            axis x line*=bottom,
            width=.19\linewidth,
            % ylabel=True label,
            xlabel=\textbf{M5},
            view={0}{90},
            y dir=reverse,
            point meta min=0,
            point meta max=1,
            xmin=-0.5,xmax=3.5,
            ymin=-0.5,ymax=3.5,
            ytick={0,1,2,3},
            xtick={0,1,2,3},
            xticklabels={},
            yticklabels={E,P,R,W},
            tick label style={font=\scriptsize},
            label style={font=\footnotesize},
            colormap={bw}{gray(0cm)=(1);gray(1cm)=(0);},
        ]
        \addplot [matrix plot*,mesh/cols=4, point meta=explicit] table [meta=z] {\confmatbittype};
        \end{axis}
        \begin{axis}[
            axis x line*=top,
            axis y line*=right,
            y dir=reverse,
            width=.19\linewidth,
            view={0}{90},   % not needed for `matrix plot*' variant
            xlabel=Predicted label,
            enlargelimits=false,
            axis on top,
            point meta min=0,
            point meta max=1,
            xmin=-0.5,xmax=3.5,
            ymin=-0.5,ymax=3.5,
            ytick={0,1,2,3},
            xtick={0,1,2,3},
            yticklabels={},
            xticklabels={E,P,R,W},
            tick label style={font=\scriptsize},
            label style={font=\footnotesize},
         ]
         \end{axis}
    \end{tikzpicture}
    %%%% ACC
   \begin{tikzpicture}
        \begin{axis}[
            enlargelimits=false,
            axis on top,
            axis x line*=bottom,
            width=.19\linewidth,
            % ylabel=True label,
            xlabel=\textbf{M6},
            view={0}{90},
            y dir=reverse,
            point meta min=0,
            point meta max=1,
            xmin=-0.5,xmax=3.5,
            ymin=-0.5,ymax=3.5,
            ytick={0,1,2,3},
            xtick={0,1,2,3},
            xticklabels={},
            yticklabels={E,P,R,W},
            tick label style={font=\scriptsize},
            label style={font=\footnotesize},
            colorbar,
            colorbar style={
                width=.02\columnwidth, 
                yticklabel style={
                    /pgf/number format/.cd,
                    fixed,
                    precision=1,
                    fixed zerofill,
                    tick label style={font=\scriptsize},
                },
            },
            colormap={bw}{gray(0cm)=(1);gray(1cm)=(0);},
        ]
        \addplot [matrix plot*,mesh/cols=4, point meta=explicit] table [meta=z] {\confmatacctype};
        \end{axis}
        \begin{axis}[
            axis x line*=top,
            axis y line*=right,
            y dir=reverse,
            width=.19\linewidth,
            view={0}{90},   % not needed for `matrix plot*' variant
            xlabel=Predicted label,
            enlargelimits=false,
            axis on top,
            point meta min=0,
            point meta max=1,
            xmin=-0.5,xmax=3.5,
            ymin=-0.5,ymax=3.5,
            ytick={0,1,2,3},
            xtick={0,1,2,3},
            yticklabels={},
            xticklabels={E,P,R,W},
            tick label style={font=\scriptsize},
            label style={font=\footnotesize},
         ]
         \end{axis}
    \end{tikzpicture}
    % \vspace{7pt}
    % \\
    % \begin{tikzpicture}
    % \begin{axis}[
    %         hide axis,  
    %         scale only axis,
    %         height=.05\columnwidth, width=0.5\columnwidth,
    %         % enlargelimits=false,
    %         % at={(frame66a.north west)}, 
    %         anchor=south west,
    %         point meta min=0,
    %         point meta max=1,
    %         colorbar horizontal,
    %         colorbar style={
    %             height=.02\columnwidth, 
    %             at={(3,1.05)},
    %             anchor=south west,
    %             xticklabel pos=lower,
    %             xticklabel style={
    %                 /pgf/number format/.cd,
    %                 fixed,
    %                 precision=1,
    %                 fixed zerofill,
    %                 tick label style={font=\scriptsize},
    %             },
    %             title={Counting normalised by true lables},
    %             title style={font=\footnotesize},
    %         },
    %         colormap={bw}{gray(0cm)=(1);gray(1cm)=(0);},
    %     ]
    %     \addplot [draw=none] coordinates {(0,0)};
    %     \end{axis}
    % \end{tikzpicture}
    \caption{Confusion matrices of filling type classification for all methods across all the containers of the public and private testing splits of the  CORSMAL Containers Manipulation dataset~\cite{Xompero_CCM}. Note that each cell is normalized by the total number of true labels for each class (colorbar). KEY -- E:~empty; P:~pasta; R:~rice; W:~water.
    \vspace{-10pt}
    }
    \label{fig:confmattype}
\end{figure*}
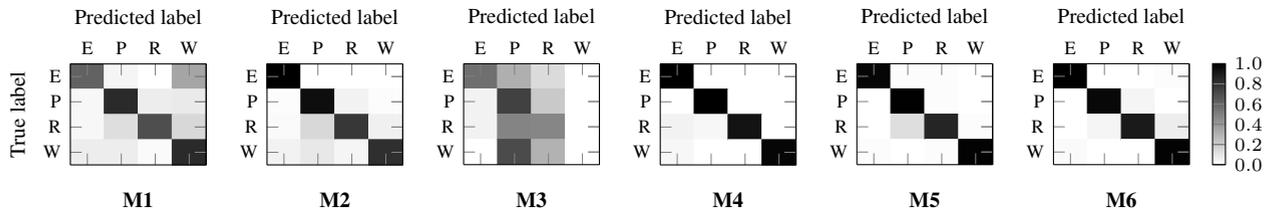

\begin{table}[t!]
    \centering
    \setlength\tabcolsep{1.5pt}
    \caption{Filling type classification  results (Task 2). Baselines and state-of-the-art methods (MX with X ranges from 1 to 6) are ranked by their score in the combined test set.
    }
    \begin{tabular}{lcccccccc}
    \specialrule{1.2pt}{0.2pt}{1pt}
    \textbf{Method} & \multicolumn{5}{c}{\textbf{Input modality}} & \multicolumn{3}{c}{\textbf{Test set}}  \\
    \cmidrule(lr){2-6}\cmidrule(lr){7-9}
     & A & R1 & R2 & R3 & R4 & Public & Private & Combined  \\
    \specialrule{1.2pt}{0.2pt}{1pt}
    Mask + RN & \wbox & \wbox & \wbox & \wbox & \bbox & 14.12 & 11.23 & 12.70 \\
    Mask + RN & \wbox & \wbox & \wbox & \bbox & \wbox & 21.14 &  9.04 & 15.63 \\
    Mask + RN & \wbox & \wbox & \bbox & \wbox & \wbox & 28.75 & 15.54 & 22.90 \\
    Mask + RN & \wbox & \bbox & \wbox & \wbox & \wbox & 30.85 & 13.04 & 23.05 \\
    Spect. + PCA + SVM  & \bbox & \wbox & \wbox & \wbox & \wbox & 20.57 & 27.60 & 24.20 \\
    \rowcolor{mylightgray}
    Random & -- & -- & -- & -- & -- & 21.24 & 27.52 & 24.38 \\
    Spect. + PCA + kNN  & \bbox & \wbox & \wbox & \wbox & \wbox & 24.47 & 28.34 & 26.53 \\
    Spect. + PCA + RF  & \bbox & \wbox & \wbox & \wbox & \wbox & 28.75 & 37.79 & 33.32 \\
    Spect. + SVM      & \bbox & \wbox & \wbox & \wbox & \wbox & 39.39 & 41.81 & 40.61 \\
    M3~\cite{Liu2020ICPR} & \bbox & \bbox & \bbox & \bbox & \bbox & 41.77 & 41.90 & 41.83 \\
    Spect. + RF       & \bbox & \wbox & \wbox & \wbox & \wbox & 47.98 & 47.68 & 47.82 \\
    Spect. + kNN      & \bbox & \wbox & \wbox & \wbox & \wbox & 60.50 & 68.58 & 64.55 \\
    AF193 + SVM         & \bbox & \wbox & \wbox & \wbox & \wbox & 64.92 & 79.72 & 72.86 \\    
    M1 & \bbox & \wbox & \wbox & \wbox & \wbox & 78.58 & 71.75 & 75.24 \\
    AF193 + kNN         & \bbox & \wbox & \wbox & \wbox & \wbox & 76.84 & 75.96 & 76.41 \\
    ZCR + MFCCs + SVM    & \bbox & \wbox & \wbox & \wbox & \wbox & 84.23 & 71.96 & 78.67 \\
    ZCR + MFCCs + kNN    & \bbox & \wbox & \wbox & \wbox & \wbox & 88.19 & 79.23 & 83.73 \\
    M2~\cite{Christmann2020NTNU} & \bbox & \wbox & \wbox & \wbox & \wbox & 81.97 & 91.67 & 86.89 \\
    AF193 + RF          & \bbox & \wbox & \wbox & \wbox & \wbox & 88.36 & 87.46 & 87.88 \\
    ZCR + MFCCs + RF     & \bbox & \wbox & \wbox & \wbox & \wbox & 92.97 & 89.74 & 91.31 \\
    M5~\cite{Iashin2020ICPR} & \bbox & \wbox & \wbox & \wbox & \wbox & 93.83 & 94.70 & 94.26 \\
    M6~\cite{Donaher2021EUSIPCO_ACC} & \bbox & \wbox & \wbox & \wbox & \wbox & 95.12 & 93.92 & 94.50 \\
    M4~\cite{Ishikawa2020ICPR} & \bbox & \wbox & \wbox & \wbox & \wbox & \textbf{97.83} & \textbf{96.08} & \textbf{96.95} \\
    \specialrule{1.2pt}{0.2pt}{1pt}
    \multicolumn{9}{l}{\scriptsize{Bold font: result of the best performing method.}}\\
    \multicolumn{9}{l}{\scriptsize{\parbox{.85\columnwidth}{KEY -- A:~audio, RX:~RGB for view X (1,2,3,4), Mask + RN:~Mask R-CNN + ResNet-18, ZCR:~zero crossing rate, MFCCs:~Mel-frequency cepstrum coefficients, Spect.:~spectrogram, RF:~random forest, SVM:~support vector machine, kNN:~k-nearest neighbor, PCA:~principal component analysis, AF193:~193 audio features consisting of MFCCs, chromogram, Mel-scaled spectrogram, spectral contrast, and tonal centroid.}}}\\
    \end{tabular}
    \label{tab:leaderboardtask2}
    % \vspace{-10pt}
\end{table}

\subsection{Filling type classification}

Table~\ref{tab:leaderboardtask2} shows that M4, M6, and M5 are the best performing methods with 96.95~$\bar{F}_1$, 94.50~$\bar{F}_1$, 94.26~$\bar{F}_1$ scores on the combined test set (as for filling level classification). Audio is the used modality by all the methods except M3 that conditions the selection of the audio-based classifier to the recognition of the container category from visual data. As for filling level classification (43.53~$\bar{F}_1$), selecting which classifier to use is likely to be the main source of error for the classifications of M3 (41.83~$\bar{F}_1$), whereas using only audio is sufficient to achieve performance close to 100~$\bar{F}_1$ score. If the audio modality was not available, both filling level and filling type classifications would be very challenging using only visual data.  M1 and M2 achieve 75.24~$\bar{F}_1$ and 86.89~$\bar{F}_1$, respectively, but about 20 and 10 percentage points (pp) lower than M4, respectively.
The table also shows that the performance of the baselines varies from random results to almost the same performance as the best performing M4. Using the spectrogram as an input feature (either after reshaping the spectrogram into a vector or after applying PCA to select the first 128 components) to any of the three classifiers, namely kNN, SVM, or RF, is the worst choice. On the combined test set, the lowest performance is obtained by Spectrogram + PCA + SVM with 24.20~$\bar{F}_1$, whereas the highest performance is obtained by Spectrogram + kNN with 64.55~$\bar{F}_1$. 
Classic audio features, such as MFCCs and ZCR, are more discriminative and sufficient to achieve performance higher than 78~$\bar{F}_1$ for the three classifiers. Simply using ZCR and MFCCs with RF can achieve 91.31~$\bar{F}_1$, which is close to the performance of the three top methods (M5, M6, M4) that are using CNNs and LSTMs. On the contrary, the performance decreases when using a larger set of features, such as tonal centroid, spectral contrast, chromogram, Mel-scaled spectrogram, and MFCCs.

Fig.~\ref{fig:confmattype} shows the confusion matrices of the methods. M4 made a few mis-classifications for the class \textit{rice} with \textit{none} and \textit{pasta}, and for the class \textit{water} with \textit{none}. M6 confused 4\% \textit{pasta} with \textit{rice}, 4\% \textit{rice} with \textit{pasta}, 7\% \textit{pasta} with \textit{water}, and 2\% \textit{water} with \textit{none}. The confusion between \textit{water} and \textit{none} could be expected due to the low volume of the sound produced by the water, whereas the confusion of \textit{water} with \textit{rice} might be caused by the glass material of the container and background noise. The largest confusion for M5 is given by the erroneous prediction of \textit{rice} with \textit{pasta} (13\%). As for filling level classification, M1 and M3 have large mis-classifications across different classes, with M3 that could not predict \textit{water} for any audio input. 

\subsection{Capacity estimation}

We compare the results of M2, M3, M4, and M5, in terms of the average capacity score. We also report the results of a pseudo-random generator (Random) that samples the predictions from a uniform distribution in the interval $[50, 4000]$ based on the Mersenne Twister algorithm~\cite{Matsumoto1998_mersenne}. We then analyze and discuss the statistics of the absolute error in predicting the container capacity for each testing container as well as for each filling type and level. 

Table~\ref{tab:leaderboardtask3} shows that M2 achieves the best score with 66.92~$\bar{C}$, 67.67~$\bar{C}$, and 67.30~$\bar{C}$ for the public test set, private test set, and the combined test set, respectively, when using only depth data from the fixed frontal view. All methods achieve a performance score that is twice higher than the random solution (24.58~$\bar{C}$ for the combined test set): M4 has the lowest score (54.79~$\bar{C}$), whereas M5 and M3 obtain 60.57~$\bar{C}$ and 62.57~$\bar{C}$, respectively. Fig.~\ref{fig:capacity_containers} shows the statistics (median, 25$^\text{th}$ and 75$^\text{th}$ quartiles, and the lower and upper whiskers\footnote{The lower whisker is the smallest data value which is larger than 0.25-quartile $-1.5\Delta$, where $\Delta$ is the difference between  0.75-quartile and 0.25-quartile. The upper whisker is the largest data value which is smaller than 0.25-quartile$+1.5\Delta$. See documentation at: \url{https://anorien.csc.warwick.ac.uk/mirrors/CTAN/graphics/pgf/contrib/pgfplots/doc/pgfplots.pdf}.}) of the relative absolute errors for each container in the test sets of the dataset. M2 has the lowest median error for all containers, except for the private containers C14 and C15. The variation of the error across configurations is either smaller than the variation of the other methods or lower than the median value of the other methods.
M5 is more consistent in estimating the same container shape and capacity for most of the configurations related to containers C12 and C15. M5 also have the largest variations for C10 and C14; M3 for C12 and C15; and M4 for C11. Interestingly, M3 have a median error lower than M4 and M5 for C13 and achieve the lowest median error with a small variation across configurations for C14. 
However, we can observe that in general the relative absolute error across containers is around or higher than 0.5. 

\begin{table}[t!]
    \centering
    \setlength\tabcolsep{2pt}
    \caption{Container capacity estimation results (Task 3). Methods ranked by the average capacity score on the combined test set.
    }
    \begin{tabular}{lccccccccccc}
    \specialrule{1.2pt}{0.2pt}{1pt}
    \textbf{Method} & \multicolumn{8}{c}{\textbf{Input modality}} & \multicolumn{3}{c}{\textbf{Test set}} \\
    \cmidrule(lr){2-9}\cmidrule(lr){10-12}
     & R1 & D1 &  R2 & D2 & R3 & D3 & R4 & D4 & Public & Private & Combined \\
    \specialrule{1.2pt}{0.2pt}{1pt}
    \rowcolor{mylightgray}
    Random      & -- & -- & -- & -- & -- & -- & -- & -- & 31.63 & 17.53 & 24.58 \\
    M4~\cite{Ishikawa2020ICPR}        & \bbox & \bbox & \wbox & \wbox & \wbox & \wbox & \wbox & \wbox & 57.19 & 52.38 & 54.79 \\
    M5~\cite{Iashin2020ICPR}         & \bbox & \wbox & \bbox & \wbox & \wbox & \wbox & \wbox & \wbox & 60.56 & 60.58 & 60.57 \\
    M3~\cite{Liu2020ICPR}        & \bbox & \wbox & \bbox & \wbox & \bbox & \wbox & \bbox & \wbox & 63.00 & 62.14 & 62.57 \\
    M2~\cite{Christmann2020NTNU}        & \wbox & \wbox & \wbox & \wbox & \wbox & \bbox & \wbox & \wbox & \textbf{66.92} & \textbf{67.67} & \textbf{67.30} \\
    \specialrule{1.2pt}{0.2pt}{1pt}
    \multicolumn{12}{l}{\scriptsize{Bold font: result of the best performing method.}}\\
    \multicolumn{12}{l}{\scriptsize{\parbox{0.9\columnwidth}{KEY -- RGB (R) or depth (D) modality for view X (1,2,3,4)}}}\\
    \end{tabular}
    \label{tab:leaderboardtask3}
    % \vspace{-10pt}
\end{table}

%%%%%%%%%%%%%%%%%%%%%%%%%%%%%%%%%%%%%%%%%%%%%%%%%%%%%%%%%%%%%%%%%%%%%%%%%%%%%%
\pgfplotsset{
    every non boxed x axis/.style={},
    boxplot/every box/.style={solid,ultra thin,black},
    boxplot/every whisker/.style={solid,ultra thin,black},
    boxplot/every median/.style={solid,very thick, red},
    title style={at={(0.5,0.98)}},
}

\pgfplotstableread{res_capacity_ntnuerc.txt}\capacityntnu
\pgfplotstableread{res_capacity_concatenation.txt}\capacityconcat
\pgfplotstableread{res_capacity_hvrl.txt}\capacityhvvrl
\pgfplotstableread{res_capacity_bit.txt}\capacitybit

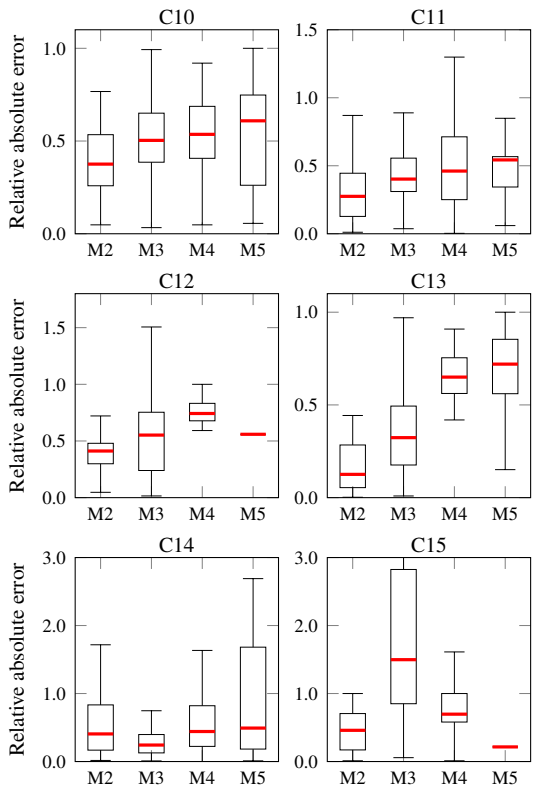
\begin{figure}[t!]
    \centering
    \begin{tikzpicture}
    \begin{axis}[
        width=.5\columnwidth,
        height=.5\columnwidth,
        xmin=0.5, xmax=4.5,
        tick label style={font=\scriptsize},
        boxplot/draw direction=y,
        xtick={1,2,3,4},
       xticklabels={M2,M3,M4,M5},
        ymin=0,ymax=1.1,
        ylabel={Relative absolute error},
        yticklabels={,0.0,0.5,1.0},
        title={C10},
        label style={font=\footnotesize},
        title style={at={(0.5,0.9)},font=\footnotesize},
    ]
    \addplot+[boxplot, boxplot/draw position=1,mark=none,boxplot/box extend=0.5] table[y=C10]{\capacityntnu};
    \addplot+[boxplot, boxplot/draw position=2,mark=none,boxplot/box extend=0.5] table[y=C10]{\capacityconcat};
    \addplot+[boxplot, boxplot/draw position=3,mark=none,boxplot/box extend=0.5] table[y=C10]{\capacityhvvrl};
    \addplot+[boxplot, boxplot/draw position=4,mark=none,boxplot/box extend=0.5] table[y=C10]{\capacitybit};
    \end{axis}
    \end{tikzpicture}
    %C11
    \begin{tikzpicture}
    \begin{axis}[
        width=.5\columnwidth,
        height=.5\columnwidth,
        xmin=0.5, xmax=4.5,
        tick label style={font=\scriptsize},
        boxplot/draw direction=y,
        xtick={1,2,3,4},
       xticklabels={M2,M3,M4,M5},
        ymin=0,ymax=1.5,
        yticklabels={,0.0,0.5,1.0,1.5},
        title={C11},
        label style={font=\footnotesize},
        title style={at={(0.5,0.9)},font=\footnotesize},
    ]
    \addplot+[boxplot, boxplot/draw position=1,mark=none, boxplot/box extend=0.5] table[y=C11]{\capacityntnu};
    \addplot+[boxplot, boxplot/draw position=2,mark=none, boxplot/box extend=0.5] table[y=C11]{\capacityconcat};
    \addplot+[boxplot, boxplot/draw position=3,mark=none, boxplot/box extend=0.5] table[y=C11]{\capacityhvvrl};
    \addplot+[boxplot, boxplot/draw position=4,mark=none, boxplot/box extend=0.5] table[y=C11]{\capacitybit};
    \end{axis}
    \end{tikzpicture}
    % C12
    \begin{tikzpicture}
    \begin{axis}[
        width=.5\columnwidth,
        height=.5\columnwidth,
        xmin=0.5, xmax=4.5,
        tick label style={font=\scriptsize},
        boxplot/draw direction=y,
        xtick={1,2,3,4},
       xticklabels={M2,M3,M4,M5},
        ymin=0,ymax=1.8,
        ylabel={Relative absolute error},
        yticklabels={,0.0,0.5,1.0,1.5},
        title={C12},
        label style={font=\footnotesize},
        title style={at={(0.5,0.9)},font=\footnotesize},
    ]
    \addplot+[boxplot, boxplot/draw position=1,mark=none,boxplot/box extend=0.5] table[y=C12]{\capacityntnu};
    \addplot+[boxplot, boxplot/draw position=2,mark=none,boxplot/box extend=0.5] table[y=C12]{\capacityconcat};
    \addplot+[boxplot, boxplot/draw position=3,mark=none,boxplot/box extend=0.5] table[y=C12]{\capacityhvvrl};
    \addplot+[boxplot, boxplot/draw position=4,mark=none,boxplot/box extend=0.5] table[y=C12]{\capacitybit};
    \end{axis}
    \end{tikzpicture}
    % C13
    \begin{tikzpicture}
    \begin{axis}[
        width=.5\columnwidth,
        height=.5\columnwidth,
        xmin=0.5, xmax=4.5,
        tick label style={font=\scriptsize},
        boxplot/draw direction=y,
        xtick={1,2,3,4},
       xticklabels={M2,M3,M4,M5},
        ymin=0,ymax=1.1,
        yticklabels={,0.0,0.5,1.0},
        title={C13},
        label style={font=\footnotesize},
        title style={at={(0.5,0.9)},font=\footnotesize},
    ]
    \addplot+[boxplot, boxplot/draw position=1,mark=none,boxplot/box extend=0.5] table[y=C13]{\capacityntnu};
    \addplot+[boxplot, boxplot/draw position=2,mark=none,boxplot/box extend=0.5] table[y=C13]{\capacityconcat};
    \addplot+[boxplot, boxplot/draw position=3,mark=none,boxplot/box extend=0.5] table[y=C13]{\capacityhvvrl};
    \addplot+[boxplot, boxplot/draw position=4,mark=none,boxplot/box extend=0.5] table[y=C13]{\capacitybit};
    \end{axis}
    \end{tikzpicture}
    % C14
    \begin{tikzpicture}
    \begin{axis}[
        width=.5\columnwidth,
        height=.5\columnwidth,
        xmin=0.5, xmax=4.5,
        tick label style={font=\scriptsize},
        boxplot/draw direction=y,
        xtick={1,2,3,4},
       xticklabels={M2,M3,M4,M5},
        ymin=0,ymax=3,
        ylabel={Relative absolute error},
        yticklabels={,0.0,1.0,2.0,3.0},
        title={C14},
        label style={font=\footnotesize},
        title style={at={(0.5,0.9)},font=\footnotesize},
    ]
    \addplot+[boxplot, boxplot/draw position=1,mark=none,boxplot/box extend=0.5] table[y=C14]{\capacityntnu};
    \addplot+[boxplot, boxplot/draw position=2,mark=none,boxplot/box extend=0.5] table[y=C14]{\capacityconcat};
    \addplot+[boxplot, boxplot/draw position=3,mark=none,boxplot/box extend=0.5] table[y=C14]{\capacityhvvrl};
    \addplot+[boxplot, boxplot/draw position=4,mark=none,boxplot/box extend=0.5] table[y=C14]{\capacitybit};
    \end{axis}
    \end{tikzpicture}
    % C15
    \begin{tikzpicture}
    \begin{axis}[
        width=.5\columnwidth,
        height=.5\columnwidth,
        xmin=0.5, xmax=4.5,
        tick label style={font=\scriptsize},
        boxplot/draw direction=y,
        xtick={1,2,3,4},
       xticklabels={M2,M3,M4,M5},
        ymin=0,ymax=3,
        yticklabels={,0.0,1.0,2.0,3.0},
        title={C15},
        label style={font=\footnotesize},
        title style={at={(0.5,0.9)},font=\footnotesize},
    ]
    \addplot+[boxplot, boxplot/draw position=1,mark=none,boxplot/box extend=0.5] table[y=C15]{\capacityntnu};
    \addplot+[boxplot, boxplot/draw position=2,mark=none,boxplot/box extend=0.5] table[y=C15]{\capacityconcat};
    \addplot+[boxplot, boxplot/draw position=3,mark=none,boxplot/box extend=0.5] table[y=C15]{\capacityhvvrl};
    \addplot+[boxplot, boxplot/draw position=4,mark=none,boxplot/box extend=0.5] table[y=C15]{\capacitybit};
    \end{axis}
    \end{tikzpicture}
    \caption{Comparison of statistics of the absolute error in estimating the container capacity for each testing container between M2~\cite{Christmann2020NTNU}, M3~\cite{Liu2020ICPR}, M4~\cite{Ishikawa2020ICPR}, and M5~\cite{Iashin2020ICPR}. Statistics of the box plot includes the median (red line), the 25\textsuperscript{th} and 75\textsuperscript{th} quartile, and the lower and upper whiskers. Note that outliers in the data are not shown. 
    Note also the different scale for the y-axis. KEY -- CX:~container (C) index (X), where X is in the range [10,15].
    }
    \label{fig:capacity_containers}
    % \vspace{-10pt}
\end{figure}
%%%%%%%%%%%%%%%%%%%%%%%%%%%%%%%%%%%%%%%%%%%%%%%%%%%%%%%%%%%%%%%%%%%%%%%%%%%%%%

In addition to the comparison across containers, Fig.~\ref{fig:capacity_fillings} shows the relative absolute errors grouped by filling type and level for each method. Most of the errors are in the interval [0.3,0.8], and the methods have similar amount of variations between the 25$^\text{th}$ and 75$^\text{th}$ quartiles, but differences are in the median error and the upper whisker error (excluding outliers). M2 achieves the lowest median error (always lower than half of the real container capacity) and smaller variations (25$^\text{th}$-75$^\text{th}$ quartiles), whereas M3 have similar results for \textit{rice full}. M4 has the largest errors for \textit{empty}, \textit{pasta half-full}, \textit{pasta full}, \textit{rice half-full}, and \textit{rice full}. M5 has the largest errors for \textit{water half-full} and \textit{water full}.

\subsection{Analysis per scenario and container}

Table~\ref{tab:scorescenariocontainer} analyzes and compares the performance scores of the methods grouped by scenario and containers for all the three tasks. For filling level classification on the testing containers, the $\bar{F}_1$ of M4, M5, and M6 increases from scenario 1 to scenario 3, showing how audio information is robust despite the increasing difficulty due to the in-hand manipulation (scenario 2 and 3) and larger distance (scenario 3). However, the performance of M6 decreases by almost 2 pp from scenario 1 (78.52~$\bar{F}_1$) to scenario 2 (76.92~$\bar{F}_1$). The performance of M1 is affected by the in-hand manipulation and distance, decreasing from 52.90~$\bar{F}_1$ in scenario 1 to 45.46~$\bar{F}_1$ in scenario 3. M3 achieves the highest accuracy for scenario 2 (51.34~$\bar{F}_1$), increasing by 11.51~pp compared to scenario 1 (39.83~$\bar{F}_1$), but decreasing to 35.92~$\bar{F}_1$ in scenario 3 (likely caused by the errors in recognizing the container category). For filling type classification, the performance of M4, M5, and M6 is higher than 90~$\bar{F}_1$ across the scenarios, but the trend is the opposite of filling level classification. M5 and M6 decrease in $\bar{F}_1$ from scenario 1 to scenario 3, whereas M4 achieves the highest accuracy in scenario 2 (98.07~$\bar{F}_1$). M3 and M1 show the same behavior for filling level and type classification with a large decrease in scenario 3 by 15.31~pp and 22.16~pp compared to scenario 1, respectively. For capacity estimation, M3 and M4 are less affected by the variations across the scenarios, whereas M2 is the best performing in scenario 1 (68.81~$\bar{C}$) and scenario 2 (73.70~$\bar{C}$) but decreases by 9.42~pp in scenario 3 compared to scenario 1. M2 is based only on the frontal depth view, where the subject is not visible for most of the time. This challenges the method to detect the object in the pre-defined depth range. M5 is affected by the increasing challenges across scenarios, decreasing from 66.51~$\bar{C}$ in scenario 1 to 55.68~$\bar{C}$ in scenario 3. This shows the limitations of the underline approach~\cite{Xompero2020ICASSP_LoDE} that was designed for objects free of occlusions and standing upright on a surface.

\pgfplotsset{
    every non boxed x axis/.style={},
    boxplot/every box/.style={solid,ultra thin,black},
    boxplot/every whisker/.style={solid,ultra thin,black},
    boxplot/every median/.style={solid,very thick, red},
    title style={at={(0.5,0.98)}},
}

\pgfplotstableread{res_capacity_ntnuerc_fillings.txt}\capacityntnuf
\pgfplotstableread{res_capacity_concatenation_fillings.txt}\capacityconcatf
\pgfplotstableread{res_capacity_hvrl_fillings.txt}\capacityhvvrlf
\pgfplotstableread{res_capacity_bit_fillings.txt}\capacitybitf

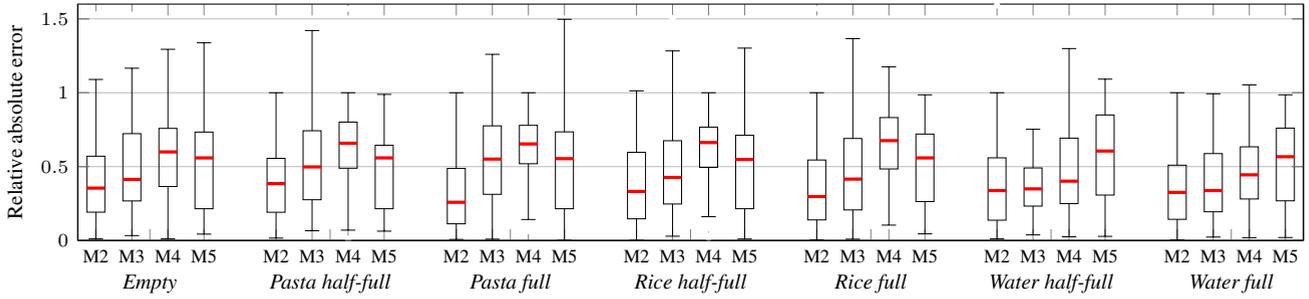
\begin{figure*}[t!]
    \centering
    \begin{tikzpicture}
    \begin{axis}[
        axis x line=box,
        width=\linewidth,
        height=.55\columnwidth,
        xmin=0.5, xmax=34.5,
        tick label style={font=\scriptsize},
        ymajorgrids=true,
        boxplot/draw direction=y,
        xtick={1,2,3,4,6,7,8,9,11,12,13,14,16,17,18,19,21,22,23,24,26,27,28,29,31,32,33,34},
       xticklabels={M2,M3,M4,M5,M2,M3,M4,M5,M2,M3,M4,M5,M2,M3,M4,M5,M2,M3,M4,M5,M2,M3,M4,M5,M2,M3,M4,M5},
        ymin=0,ymax=1.6,
        ylabel={Relative absolute error},
        label style={font=\footnotesize},
        title style={font=\footnotesize},
    ]
    \addplot+[boxplot, boxplot/draw position=1,mark=*, mark options={white,scale=0.5},boxplot/box extend=0.5] table[y=E]{\capacityntnuf};
    \addplot+[boxplot, boxplot/draw position=2,mark=*, mark options={white,scale=0.5},boxplot/box extend=0.5] table[y=E]{\capacityconcatf};
    \addplot+[boxplot, boxplot/draw position=3,mark=*, mark options={white,scale=0.5},boxplot/box extend=0.5] table[y=E]{\capacityhvvrlf};
    \addplot+[boxplot, boxplot/draw position=4,mark=*, mark options={white,scale=0.5},boxplot/box extend=0.5] table[y=E]{\capacitybitf};
    \addplot+[boxplot, boxplot/draw position=6,mark=*, mark options={white,scale=0.5},boxplot/box extend=0.5] table[y=P5]{\capacityntnuf};
    \addplot+[boxplot, boxplot/draw position=7,mark=*, mark options={white,scale=0.5},boxplot/box extend=0.5] table[y=P5]{\capacityconcatf};
    \addplot+[boxplot, boxplot/draw position=8,mark=*, mark options={white,scale=0.5},boxplot/box extend=0.5] table[y=P5]{\capacityhvvrlf};
    \addplot+[boxplot, boxplot/draw position=9,mark=*, mark options={white,scale=0.5},boxplot/box extend=0.5] table[y=P5]{\capacitybitf};
    \addplot+[boxplot, boxplot/draw position=11,mark=*, mark options={white,scale=0.5},boxplot/box extend=0.5] table[y=P9]{\capacityntnuf};
    \addplot+[boxplot, boxplot/draw position=12,mark=*, mark options={white,scale=0.5},boxplot/box extend=0.5] table[y=P9]{\capacityconcatf};
    \addplot+[boxplot, boxplot/draw position=13,mark=*, mark options={white,scale=0.5},boxplot/box extend=0.5] table[y=P9]{\capacityhvvrlf};
    \addplot+[boxplot, boxplot/draw position=14,mark=*, mark options={white,scale=0.5},boxplot/box extend=0.5] table[y=P9]{\capacitybitf};
    \addplot+[boxplot, boxplot/draw position=16,mark=*, mark options={white,scale=0.5},boxplot/box extend=0.5] table[y=R5]{\capacityntnuf};
    \addplot+[boxplot, boxplot/draw position=17,mark=*, mark options={white,scale=0.5},boxplot/box extend=0.5] table[y=R5]{\capacityconcatf};
    \addplot+[boxplot, boxplot/draw position=18,mark=*, mark options={white,scale=0.5},boxplot/box extend=0.5] table[y=R5]{\capacityhvvrlf};
    \addplot+[boxplot, boxplot/draw position=19,mark=*, mark options={white,scale=0.5},boxplot/box extend=0.5] table[y=R5]{\capacitybitf};
    \addplot+[boxplot, boxplot/draw position=21,mark=*, mark options={white,scale=0.5},boxplot/box extend=0.5] table[y=R9]{\capacityntnuf};
    \addplot+[boxplot, boxplot/draw position=22,mark=*, mark options={white,scale=0.5},boxplot/box extend=0.5] table[y=R9]{\capacityconcatf};
    \addplot+[boxplot, boxplot/draw position=23,mark=*, mark options={white,scale=0.5},boxplot/box extend=0.5] table[y=R9]{\capacityhvvrlf};
    \addplot+[boxplot, boxplot/draw position=24,mark=*, mark options={white,scale=0.5},boxplot/box extend=0.5] table[y=R9]{\capacitybitf};
    \addplot+[boxplot, boxplot/draw position=26,mark=*, mark options={white,scale=0.5},boxplot/box extend=0.5] table[y=W5]{\capacityntnuf};
    \addplot+[boxplot, boxplot/draw position=27,mark=*, mark options={white,scale=0.5},boxplot/box extend=0.5] table[y=W5]{\capacityconcatf};
    \addplot+[boxplot, boxplot/draw position=28,mark=*, mark options={white,scale=0.5},boxplot/box extend=0.5] table[y=W5]{\capacityhvvrlf};
    \addplot+[boxplot, boxplot/draw position=29,mark=*, mark options={white,scale=0.5},boxplot/box extend=0.5] table[y=W5]{\capacitybitf};
    \addplot+[boxplot, boxplot/draw position=31,mark=*, mark options={white,scale=0.5},boxplot/box extend=0.5] table[y=W9]{\capacityntnuf};
    \addplot+[boxplot, boxplot/draw position=32,mark=*, mark options={white,scale=0.5},boxplot/box extend=0.5] table[y=W9]{\capacityconcatf};
    \addplot+[boxplot, boxplot/draw position=33,mark=*, mark options={white,scale=0.5},boxplot/box extend=0.5] table[y=W9]{\capacityhvvrlf};
    \addplot+[boxplot, boxplot/draw position=34,mark=*, mark options={white,scale=0.5},boxplot/box extend=0.5] table[y=W9]{\capacitybitf};
    \end{axis}
    \begin{axis}[
        axis x line=bottom,
        % axis y line=none,
        width=\linewidth,
        height=.55\columnwidth,
        xmin=0.5, xmax=34.5,
        xticklabel shift={8},
        tick label style={font=\footnotesize},
        xtick={2.5,7.5,12.5,17.5,22.5,27.5,32.5},
        xticklabels={\textit{Empty},\textit{Pasta half-full}, \textit{Pasta full},\textit{Rice half-full}, \textit{Rice full},\textit{Water half-full}, \textit{Water full}},
        typeset ticklabels with strut,
        ymin=0,ymax=1.6,
        yticklabels={},
        label style={font=\footnotesize},
    ]
    \end{axis}
    \end{tikzpicture}
    \vspace{-10pt}
    \caption{Comparison of the absolute error in estimating the container capacity between M2~\cite{Christmann2020NTNU}, M3~\cite{Liu2020ICPR}, M4~\cite{Ishikawa2020ICPR}, and M5~\cite{Iashin2020ICPR} for the different combinations of filling type and level in the combined public and private test set of the CORSMAL Containers Manipulation dataset. Statistics of the box plot includes the median (red line), the 25\textsuperscript{th} and 75\textsuperscript{th} percentiles, and the minimum and maximum error..
    %Note that limit the y-axis to 1,000 as the most significant results are below this value.
    }
    \label{fig:capacity_fillings}
    \vspace{-10pt}
\end{figure*}

\begin{table}[t!]
    \centering
    \footnotesize
    \setlength\tabcolsep{1.5pt}
    \caption{Comparison of the task performance scores between methods for each scenario and for each testing container.}
    \begin{tabular}{llcccrrrrrr}
    \specialrule{1.2pt}{0.2pt}{1pt}
     & \textbf{Method} & \textbf{S1} & \textbf{S2} & \textbf{S3} & \textbf{C10} & \textbf{C11} & \textbf{C12} & \textbf{C13} & \textbf{C14} & \textbf{C15}  \\
    \specialrule{1.2pt}{0.2pt}{1pt}
    \multirow{5}{*}{T1} 
    & M1             & 52.90 & 47.37 & 45.46 & 48.69 & 46.78 & 58.27 & 41.43 & 38.89 & 62.58 \\
    & M3~\cite{Liu2020ICPR}                 & 39.83 & 51.34 & 35.92 & 35.24 & 36.59 & 22.86 & 33.74 & 33.74 & 26.33 \\
    & M4~\cite{Ishikawa2020ICPR}                    & 75.41 & 77.70 & 82.54 & 92.85 & 89.25 & 46.67 & 86.85 & 74.16 & 45.92 \\
    & M5~\cite{Iashin2020ICPR}                     & 75.87 & 80.89 & 82.03 & 83.12 & 90.48 & 41.26 & 88.09 & 90.36 & 47.68 \\
    & M6~\cite{Donaher2021EUSIPCO_ACC}                     & 78.52 & 76.92 & 86.84 & 83.12 & 88.10 & 64.98 & 89.16 & 78.28 & 74.99 \\
    \midrule
    \multirow{6}{*}{T2} 
    & M1             & 81.22 & 77.01 & 66.06 & 86.42 & 69.33 & 79.67 &  87.38 & 55.60 & 72.77 \\
    & M2~\cite{Christmann2020NTNU}                & 90.68 & 84.57 & 85.41 & 77.29 & 80.60 & 91.58 &  94.02 & 94.09 & 85.03 \\
    & M3~\cite{Liu2020ICPR}                 & 44.06 & 51.21 & 28.75 & 21.72 & 26.54 & 86.98 &  20.33 & 34.13 & 79.45 \\
    & M4~\cite{Ishikawa2020ICPR}                    & 97.35 & 98.07 & 95.45 & 97.63 & 98.82 & 96.72 & 100.00 & 97.66 & 87.96 \\
    & M5~\cite{Iashin2020ICPR}                     & 96.70 & 94.76 & 91.32 & 96.44 & 97.62 & 84.81 &  97.63 & 98.81 & 84.45 \\
    & M6~\cite{Donaher2021EUSIPCO_ACC}                     & 96.70 & 95.43 & 91.27 & 91.58 & 96.41 & 98.33 & 97.62 & 85.61 & 100.00 \\
    \midrule
    \multirow{4}{*}{T3} 
    & M2~\cite{Christmann2020NTNU}              & 68.81 & 73.70 & 59.39 & 66.02 & 69.14 & 65.08 & 79.75 & 61.12 & 59.94 \\
    & M3~\cite{Liu2020ICPR}                 & 64.33 & 60.41 & 62.96 & 60.99 & 66.21 & 61.30 & 71.90 & 76.75 & 28.02 \\
    & M4~\cite{Ishikawa2020ICPR}                 & 55.45 & 55.34 & 53.57 & 59.62 & 61.70 & 47.47 & 53.77 & 58.29 & 42.17 \\
    & M5~\cite{Iashin2020ICPR}                    & 66.51 & 59.51 & 55.68 & 60.71 & 62.43 & 57.71 & 53.37 & 54.75 & 78.82 \\
    \specialrule{1.2pt}{0.2pt}{1pt}
    \multicolumn{11}{l}{\scriptsize{KEY -- S:~scenario, C:~container, T:~task.}}
    \end{tabular}
    \label{tab:scorescenariocontainer}
    % \vspace{-10pt}
\end{table}

The performance across containers varies between the methods. Testing containers 12 and 15 are the most challenging for M3, M4, M5, M6, when classifying the filling level, whereas M1 achieves its best performance on both containers. M4 and M5 have the largest decrease with the score in the interval [40,50]~$\bar{F}_1$ compared to the interval [75-93]~$\bar{F}_1$ for the other containers. M6 outperforms all the other methods with 64.98~$\bar{F}_1$ and 74.99~$\bar{F}_1$ for containers 12 and 15. For filling type classification, M3 obtains 86.98~$\bar{F}_1$ and 79.45~$\bar{F}_1$ for containers 12 and 15, respectively, and less than 30~$\bar{F}_1$ on the other containers. Because of the dataset structure, M3 can recognize the \textit{box} class and the filling type for that class, but the method cannot easily distinguish filling type and level for drinking glasses and cups. Overall, other methods achieve a score higher than 70~$\bar{F}_1$ across containers. M4 achieves 100~$\bar{F}_1$ on container 13 and M6 on container 15. M4 is the best performing for containers 10 and 11, whereas M5 is the best for container 14. Containers 12 and 15 are the most challenging for M5; container 14 for M6; container 15 for M4; containers 10, 11, and 15 for M2. M1 ranges between 55.60~$\bar{F}_1$ and 87.38~$\bar{F}_1$ across containers, with the drinking glasses being the most challenging and obtaining 69.33~$\bar{F}_1$ for container 11 and 55.60~$\bar{F}_1$ for container 14. For capacity estimation, M2 achieves the best performance on containers 10 (66.02~$\bar{C}$), 11 (69.14~$\bar{C}$), 12 (65.02~$\bar{C}$), and 13 (79.75~$\bar{C}$), M3 on container 14 (76.75~$\bar{C}$), and M5 on container 15 (78.82~$\bar{C}$). M3 achieves higher average capacity score on the private cup and drinking glass than the public containers, but the score drops to 28.02~$\bar{C}$ for the container 15. M4 performs worse on the private testing containers than the public testing containers, with the lowest scores on the boxes (containers 12 and 15). M5 also performs worse for the drinking glass and cups in the private test set than the public test set. Surprisingly, the best score of M5 is on the \textit{box} container 15 (78.82~$\bar{C}$) despite the modeled shape is a 3D cylinder.

\begin{table}[t!]
    \centering
    \footnotesize
    \caption{Comparison of the filling mass estimation results. Methods are ranked by their score on the combined test sets of the CORSMAL Containers Manipulation dataset. Note that scores are weighed by the number of tasks addressed by the methods.}
    \begin{tabular}{lcccccc}
    \specialrule{1.2pt}{0.2pt}{1pt}
    \textbf{Method} & \multicolumn{3}{c}{\textbf{Task}} & \multicolumn{3}{c}{\textbf{Test set}}  \\
    \cmidrule(lr){2-4}\cmidrule(lr){5-7}
     & T1 & T2 & T3 & Public & Private & Combined  \\
    \specialrule{1.2pt}{0.2pt}{1pt}
    M6~\cite{Donaher2021EUSIPCO_ACC}        & \bbox & \bbox & \wbox & 28.25 & 21.89 & 25.07 \\
    M1    & \bbox & \bbox & \wbox & 29.25 & 23.21 & 26.23 \\
    \rowcolor{mylightgray}Random & \bbox & \bbox & \bbox & 38.47 & 31.65 & 35.06  \\
    M2~\cite{Christmann2020NTNU}       & \wbox & \bbox & \bbox & 38.56 & 39.80 & 39.18 \\
    M3~\cite{Liu2020ICPR}         & \bbox & \bbox & \bbox & 52.80 & 51.14 & 53.47 \\
    M4~\cite{Ishikawa2020ICPR}        & \bbox & \bbox & \bbox & 63.32 & 61.01 & 62.16 \\
    M5~\cite{Iashin2020ICPR} & \bbox & \bbox & \bbox & \textbf{64.98} & \textbf{65.15} & \textbf{65.06} \\
    \specialrule{1.2pt}{0.2pt}{1pt}
    \end{tabular}
    \label{tab:leaderboardmass}
    %\vspace{-10pt}
\end{table}

\subsection{Filling mass estimation}

We discuss the overall performance of the methods based on their results on estimating the filling mass. Methods that estimated either of the physical properties in our framework (e.g., M1, M2, and M6) are complemented by the random estimation of the missing physical properties to compute the filling  mass\footnote{Note that for the organized challenge, the score is weighted by the number of completed tasks. We report the results in the same manner.}. Table~\ref{tab:leaderboardmass} shows that methods addressing only filling type and level classification achieve a lower score than a random guess for each task. Given the multiplicative formula of the filling mass estimation (see Eq.~\ref{eq:fillingmass}), even a few errors in these classification tasks can lead to a low score in the filling mass estimation, especially when combined with the random estimation of the container capacity. However, improving the capacity estimation is an important aspect to achieve more accurate results (and higher score) for the filling mass estimation (see M2). M3, M4, and M5 addressed all three tasks and achieved 53.47~$\bar{M}$, 62.16~$\bar{M}$, and 65.06~$\bar{M}$, respectively. Overall, methods perform better on the public test set than the private test set, except for M2 and M5 that achieve similar performance in the two test sets. We can observe that the more accurate predictions in the container capacity help M3 to obtain 53.47~$\bar{M}$ despite the classification errors for filling level and type. The high classification accuracy on filling level and type, combined with a similar score for the capacity estimations with respect to M3, makes M4 and M5 the best performing in filling mass estimation. The similar scores for container capacity and filling mass estimation shows how important it is to accurately predict the capacity in order to correctly estimate the filling mass.

\section{Conclusion}
\label{sec:conclusions}

We presented the open CORSMAL framework to benchmark methods for estimating the physical properties of different containers while they are manipulated by a person with different content types. The framework includes a dataset, a set of tasks and performance measures, and several baselines that use either audio or visual input. The framework supports the contactless estimation of the weight of the container, including its content (if any), despite variations in the physical properties across containers and occlusions caused by the hand manipulation. 

We performed an in-depth comparative analysis of the baselines and state-of-the-art methods that used the framework. The analysis showed that using only audio as input is sufficient to achieve a weighted average F1-score above 80\% for filling type and level classification, but the high performance could be limited to the sensor types and setup of the CORSMAL Container Manipulation dataset. Methods that use audio alone are robust to changes in the container type, size, and shape, as well as pose during the manipulation. Moreover, filling type and level estimation can benefit from each other to avoid unfeasible solutions~\cite{Donaher2021EUSIPCO_ACC}. Container capacity is the most challenging physical property to estimate with all methods affected by large errors and a maximum score of 65\%. Performance on this task also affects the successive estimation of the filling mass. The design of a method that can generalize across the different containers and scenarios, especially for container capacity estimation and partially for filling level classification, is still challenging. 

Future directions involve the exploration of fusion and learning methods with both acoustic and visual modalities to support the contactless estimation of the physical properties of containers and their content. The CORSMAL framework is open for further submissions and support the research in this upcoming area\footnote{\url{https://corsmal.eecs.qmul.ac.uk/challenge.html}}.

\section*{Acknowledgment}
We would like to thank Ricardo Sanchez-Matilla and Riccardo Mazzon for their contribution in the design and collection of the data, and the performance measures definition.

\bibliographystyle{IEEEtran}
\bibliography{main}

\EOD

\end{document}